\documentstyle[epsf]{mn}

\newif\ifAMStwofonts
\AMStwofontstrue

\title[FIR Survey of Molecular Cloud Cores]
{A Far-Infrared Survey of Molecular Cloud Cores}
\author[N. E. Jessop and D. Ward-Thompson]
       {N. E. Jessop$^1$ and D. Ward-Thompson$^2$ \\
       $^1$Observatorio Astronomico de Lisboa, Universidade de Lisboa,
Portugal \\
       $^2$Department of Physics and Astronomy, University of Wales, 
Cardiff}
\date{Accepted 1999 April 1; Received 1999 March 1.}

\pagerange{\pageref{firstpage}--\pageref{lastpage}}
\pubyear{1999}

\begin{document}

\maketitle

\label{firstpage}

\begin{abstract}
We present a catalogue of molecular cloud cores drawn from high latitude, 
medium opacity clouds, using the all-sky IRAS Sky Survey Atlas (ISSA)
images at 60 and 100~$\mu$m. The typical column densities of the cores are
$ \rm N(H_2)\sim 3.8 \times 10^{21} $cm$\rm ^{-2}$ and the typical volume
densities are $ \rm n(H_2) \sim 2 \times 10^3$cm$\rm ^{-3}$.
They are therefore significantly less dense than many other samples
obtained in other ways. Those cloud cores with {\it IRAS} point
sources are seen to be already forming stars, but this is found to be only
a small fraction of the total number of cores. The fraction of the
cores in the protostellar stage is used to estimate the
prestellar timescale -- the time until the formation of a
hydrostatically supported
protostellar object. We argue, on the basis of a comparison 
with other samples, that a trend exists
for the prestellar lifetime of a cloud core to decrease with the
mean column density and number density of the core.
We compare this with model predictions and show that 
the data are consistent with
star formation regulated by the ionisation fraction.
\end{abstract}


\section{Introduction}

The denser regions of the interstellar medium are usually known as
molecular clouds, and stars form in the very
highest density parts of these clouds, which are usually
referred to as molecular cloud cores. 
The process by which material is turned from a molecular cloud core into a
star is far from being fully understood, since many complex physical
processes are involved (see, for example, Mouschovias 1991, for a review).

The key to understanding the process of star formation appears to lie in
the earliest stages of molecular cloud core evolution, since the initial
conditions determine much of what subsequently takes place.
Many observational studies have
been conducted of such dense cores (e.g. Myers \& Benson 1983;
Myers, Linke \& Benson 1983;
Clemens \& Barvainis 1988; Benson \& Myers 1989; Bourke, Hyland \& 
Robinson 1995a), usually starting from an optically selected sample 
of dark clouds on sky survey plates. In this paper we 
present a sample of cloud cores selected on the basis of far-infrared
optical depth using the IRAS Sky Survey Atlas (ISSA) images (Wheelock
et al. 1994), with a view to broadening the range of the physical
parameters of cores that have been explored, and hence gaining further
insight into the evolution of molecular cloud cores.

The paper is laid out as follows: Section 2 describes the manner in which
optical depth maps were constructed from the ISSA data, including
technical details such as background subtraction, and goes on to discuss
cloud core selection and methodology
verification techniques; Section 3 describes the new molecular cloud core
catalogue that we have constructed from the data and the associations with
previously known molecular clouds and IRAS point sources; Section 4
contrasts the properties of the new catalogue with previous molecular cloud
core catalogues and compares the ensemble of mean catalogue properties
with the theoretical predictions of models of ionisation-regulated star
formation; and Section 5 presents the conclusions of the paper. Readers
who are more theoretically inclined may wish to read Sections 3 \& 4
before going back to Section 2.

\section{Observational details}

\subsection{Background subtraction}

We chose to select the clouds using the {\it IRAS} Sky Survey Atlas (ISSA),
which is an all-sky set of images at each of the four IRAS wavebands of 
12, 25, 60 \& 100~$\mu$m (Beichman et al. 1988;
Wheelock et al. 1994). However,
since we wished to select the coldest, densest clouds, we concentrated
on the two longest wavelengths, 60 \& 100~$\mu$m. We used these to construct
optical depth and temperature maps of the regions of interest, in a 
similar manner to that adopted by Wood, Myers \& Daugherty (1994). 
These latter authors selected previously known molecular cloud regions,
whereas we endeavoured to construct a 
distinct sample of previously little-studied
clouds drawn from the all-sky set of plates.
But before optical depth and temperature maps can be made,
it is necessary to ensure that all background emission has been removed from
the images. This is because in making optical depth and temperature maps,
one must take the ratio of emission at two different wavelengths (in this
case 60 \& 100~$\mu$m). Therefore, any offset at one or other wavelength
due to background emission not associated with the object will affect
the ratio measurement. The ISSA images have already been fully processed
and corrected for most of the
Zodiacal background emission, but they still contain 
extended emission from the Galactic Plane and some residual Zodiacal
emission. These are the backgrounds we now address.

Pixel histograms of all the ISSA fields were 
constructed by binning the pixels in each field into histograms of 
surface intensity. 
On examination it was found that some fields' pixel histograms contained a 
single peak, some contained double or multiple peaks and some had much more 
complicated structures. Each field was 
searched to identify whether the histogram had 
a peak with a width of less than 
2MJysr$^{-1}$ (i.e. $ \rm \sim$ 10 $ \rm \times$ the calibration noise) and at
low enough intensity to be consistent with an area of background. 
Any such 
peak was taken as evidence of an area of low level emission in the field 
which can be described as an area of background containing only low level 
cirrus. A number of the fields were inspected visually. It was found that the 
pixels in the low level peak of the histogram in each case
came from a contiguous, discrete area of the sky, and were not randomly 
isolated pixels, confirming that this automatic method was indeed finding
genuine areas of sky.

A similar method was previously carried out
for optical images by Beard, MacGillivray \& 
Thanisch (1990), who showed that histograms of pixels' intensities contain 
valuable information about the background 
regions within the image. In relatively empty areas of the sky the pixel 
histogram takes the form of a single Gaussian peak whereas in regions densely 
populated with real sources several further peaks appear in the histogram at 
higher intensities. 

The position of the peak was recorded and used as a measure of the
background surface brightness in the field. 
272 ISSA fields were identified as having suitable background regions
by this technique at 100$ \rm \mu$m and 368 were found at 
60$ \rm \mu$m from a total of 430 for each wavelength. The 
widths of the low intensity peaks in the pixel histograms 
were generally at least 
a factor of two or three larger 
than the average calibration errors 
for the field -- as estimated from the standard deviation in
pixel values from one Hours Confirmed scan (HCON) to 
the next -- indicating the existence of real cirrus structure within the
background regions.

To investigate the overall nature of the background, 
maps which interpolated from region to region of background were constructed.
This was done by first recording the values of the pixel histogram low 
intensity peaks (in MJysr$^{-1}$) and recording the central position of the 
field in which it was found.
A simple interpolation technique was devised, 
somewhat akin to box-car averaging.
For every square degree on the celestial sphere 
the distance to 
the centre of the nearest few ISSA fields containing a region of 
background was calculated, to allow averaging. 
This gives a series of angular displacements 
to the background regions 
($  \theta_{\rm{1}},\theta_2,\theta_3,\theta_4,....\theta_n$),
where:

\[ \rm{cos}(\theta_{\rm{i}})=\rm{cos(dec_{i}) cos(dec) cos(ra-ra_i) -
sin(dec_i)sin(dec)}. \]

\noindent
Here the right ascensions and declinations of the positions to which
we wish to interpolate are denoted ra and dec 
respectively, and the position of each background region is denoted 
$ \rm ra_i$ and $ \rm dec_i$.
We then attributed to each position the value of surface brightness:

\[ I = \frac{ \sum^{\theta_{\rm{i}} <\phi} 
{\rm cos}\left( \frac{\pi\theta_{{\rm i} }}{\phi}\right)I_{{\rm i}}} 
{\sum^{\theta_{{\rm i}} <\phi} \rm{cos}\left( \frac{\pi\theta_{{\rm i}} }
{\phi}\right)}. \]

\noindent
This was chosen because it interpolates between positions smoothly,
does not produce discontinuities and weights more heavily the nearest 
background regions, even given the non-uniform spatial sampling of 
background surface brightness.  It essentially involves smoothing 
by a cosine bell of radius $ \phi$. The result of this 
process was then projected onto a 2d surface with equal area 
projection to create a map of background brightness for the whole sky.

Fig. 1 shows the map of the background constructed by this technique
at 100$\mu$m for $ \phi =$ 12 degrees. The plot is an `equal area' 
projection in Galactic coordinates. The intensity varies smoothly,
and increases towards the Galactic Plane 
indicating that the cirrus intensity
within background regions increases towards the Plane. The map
contains negative values showing that the Zodiacal
background subtraction used in producing the ISSA 
dataset was liable to over-compensate in some regions. 
The apparent gap through the centre of the map is due to our having to
discard regions of high source confusion in the Galactic Plane, where
our Galactic background subtraction did not work due to source confusion.

\begin{figure}
\begin{center}
\epsfxsize=8cm
\epsfbox[50 30 470 280]{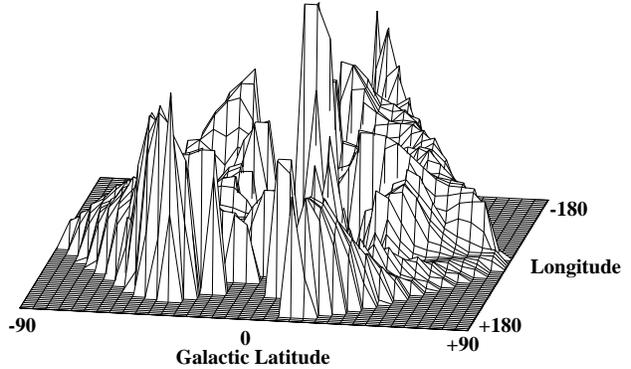}
\caption{Interpolated map of 100-$\mu$m background regions. 
Values range from 
$-$0.2 to $+$11 MJySr$^{-1}$. There is a gap in the data
through the map along the Galactic Plane 
(partly obscured because of the projection), due to a larger 
number density of clouds near the Galactic plane
causing source confusion and thus preventing the background subtraction
method from working.}
\end{center}
\end{figure}

The residual `striping' characteristic of images produced by IRAS was
reduced to very low levels in the ISSA images, because of the careful
calibration used to produce them. However, the act of ratio-ing two
images tends to enhance any striping effects that remain. In some fields
this striping was clearly visible, and in some cases dominated the
structure of the optical depths maps. These fields were discarded.
We selected 60 of the fields used in the interpolation for further study,
based on their background regions being clearly identifiable and
measurable, and on the quality of the images.

\begin{center}
\begin{table*}
\caption{List of cloud positions, associations, velocities and distances.
In column 4 a superscript indicates that the association is a:
(1) dark nebula catalogued 
by Lynds (1962); (2) bright nebula
catalogued by Lynds (1965); (3) molecular 
cloud identified and mapped in CO by Taylor, Dickman \& Scoville (1987); 
(4) cluster listed by Lang 
(1992); and (5) molecular cloud 
identified and mapped in CO by Ramesh (1994).}
\begin{tabular}{cccccc}
\hline
Cloud & R.A. & Dec. & Associations & Velocity & Distance \\ 
Name  & (1950) & (1950) & identified & (LSR)/kms$^{-1}$ & (pc) \\ \hline
 021A & 00$^{\rm h}$ 00$^{\rm m}$ & $-$76$^{\circ}$ 00$^{\prime}$ & 
Chameleon & -- & 200 \\
 270A & 00$^{\rm h}$ 02$^{\rm m}$ & $+$13$^{\circ}$ 50$^{\prime}$ & 
None & -- & $ \rm <$80 \\ 
 422A & 01$^{\rm h}$ 30$^{\rm m}$ & $+$76$^{\circ}$ 30$^{\prime}$ & 
Cepheus  & -- & 300--800 \\
 423A & 02$^{\rm h}$ 10$^{\rm m}$ & $+$75$^{\circ}$ 30$^{\prime}$ & 
Cepheus, L133$\rm ^1$, LBN$ \rm ^2$, & $+$3.1$\rightarrow +$3.5 & 800 \\
 & & &  T486$ \rm ^3$, T486$ \rm ^4$, Berkeley8$ \rm ^4$ &  & \\
 002B & 02$^{\rm h}$ 30$^{\rm m}$ & $-$85$^{\circ}$ 00$^{\prime}$ & 
Chameleon & -- & 200 \\
 423B & 03$^{\rm h}$ 00$^{\rm m}$ & $+$81$^{\circ}$ 20$^{\prime}$ & 
Cepheus & -- & 300--800 \\
 411A & 04$^{\rm h}$ 15$^{\rm m}$ & $+$64$^{\circ}$ 20$^{\prime}$ & 
Cepheus & -- & 300--800 \\
 205A & 04$^{\rm h}$ 28$^{\rm m}$ & $+$04$^{\circ}$ 00$^{\prime}$ & 
Orion & -- & 450 \\
 050A & 07$^{\rm h}$ 26$^{\rm m}$ & $-$48$^{\circ}$ 00$^{\prime}$ & 
R42$ \rm ^5$, Merlotte 66$ \rm ^4$ & $+$4.8 & $ \rm <$250 \\
 221A & 15$^{\rm h}$ 44$^{\rm m}$ & $-$04$^{\circ}$ 00$^{\prime}$ & 
$ \rm \rho$ Ophiuchus, L134N$ \rm ^1$ & $+$2.6 & 125 \\
 420C & 20$^{\rm h}$ 40$^{\rm m}$ & $+$67$^{\circ}$ 00$^{\prime}$ & 
Cepheus, L1148$ \rm ^1$, T379$ \rm ^3$ & $+$3.5 & 300 \\
 420B & 21$^{\rm h}$ 30$^{\rm m}$ & $+$66$^{\circ}$ 00$^{\prime}$ & 
Cepheus, L1176$ \rm ^1$, T309$ \rm ^3$, & $-$10.8 & 800 \\ 
 & & & NGC 7142$ \rm ^4$, NGC 7129$ \rm ^4$ &  & \\
 002A & 21$^{\rm h}$ 30$^{\rm m}$ & $-$83$^{\circ}$ 00$^{\prime}$ & 
Chameleon & -- & 200 \\
 334A & 21$^{\rm h}$ 48$^{\rm m}$ & $+$35$^{\circ}$ 30$^{\prime}$ & 
 None & -- & $ \rm <$270 \\
 422C & 22$^{\rm h}$ 00$^{\rm m}$ & $+$77$^{\circ}$ 30$^{\prime}$ & 
Cepheus, LBN$^2$, & $-$1.7$\rightarrow +$5.1 & 300 \\
 & & & T420$ \rm ^3$,T428$ \rm ^3$,T431$ \rm ^3$  & & \\ 
 422B & 22$^{\rm h}$ 45$^{\rm m}$ & $+$73$^{\circ}$ 40$^{\prime}$ & 
Cepheus, T426$ \rm ^3$, T431$ \rm ^3$ & $-$3.7 & 300 \\ \hline
\end{tabular}
\end{table*}
\end{center}

\subsection{Colour temperatures}

We used the STARLINK data reduction package IRAS90 and, in particular the
routine COLTEMP, to create colour temperature and
optical depth maps from the ISSA images. We briefly
describe the technique here (for further details, see Berry 1993a \& b).

The far infrared radiation from a cloud of temperature 
$ \rm T $ emitting a
black body spectrum, $ \rm B(\nu,T) $, and absorbing 
$ \rm I(\nu,T)d\tau_\nu$, at a position in the cloud with an optical depth 
$ \rm \tau_\nu$,
leads to a flux received by an observer $ \rm f(\nu,T)$ given by:

\[ \rm f(\nu,T)=\left(1-e^{-\tau_{\nu}}\right)B(\nu,T). \]

Generally $ \rm \tau_\nu$ is dependent on $ \rm \nu$ in such a way
(Hildebrand 1983) that:

\[\rm
\tau(\nu) = \left( \frac{\nu}{\nu_c}\right)^{\beta}, \]

\noindent
where $ \rm \beta$ is the dust emissivity index and $\nu_c$ is the critical 
frequency at which the optical depth is unity.
Throughout this
work we use the value suggested by Hildebrand (1983) of 
$\rm \beta =1$ at far-infrared wavelengths.
These two equations describe a `greybody' spectrum.
We assume that the cloud is optically thin at the wavelengths observed
-- Wood et al. (1994) show that even towards the Galactic Plane this is true.
Because
$ \rm \tau_\nu \ll 1$, $ \rm 1-e^{-\tau_\nu } \sim \tau_\nu $  and we 
therefore use the expression:

\[ \rm f(\nu,T) \sim \left( \frac{\nu}{\nu_c}\right)B(\nu,T). \]

The IRAS detectors were sensitive over a wide bandpass and there were 4 
separate wavebands (i=1, 2, 3, 4 for 12, 25, 60 and 100 $ \rm \mu$m 
respectively) 
so that the 
measured flux in waveband i is:

\[\rm f_i= \int R_i(\nu)\left( \frac{\nu}{\nu_c}\right)B(\nu,T)d\nu \]

\noindent
where $ \rm R_i(\nu)$ is the spectral response curve for the waveband 
i receiver (see Beichman et al. 1988).

By taking the ratio of intensities at two wavebands i and j, and using 
the preceding equation, one can derive 

\[\rm
\frac{f_i}{f_j} = \frac{\int R_i(\nu) \nu B(\nu,T) d\nu}
{\int R_j(\nu)\nu B(\nu,T) d\nu}. \]

This value is dependent on T, and the response curves
of the receivers. Using the listed
response curves in the IRAS Explanatory Supplement (Beichman et al. 1988),
one can tabulate $ \rm f_i/f_j$ versus T. In the routine COLTEMP a spline 
giving T as a function of $ \rm f_i/f_j$ is created by fitting to the 
tabulated values. For any observations of a cloud at 2 wavelengths one can 
then estimate the temperature and calculate the critical frequency at
which $ \rm \tau_{\nu_c} = 1$:

\[ \rm
\nu_{c} = \frac{\int R_i(\nu) \nu B(\nu,T)d \nu}{f_i}. \]

Using this, one can calculate the optical depth of the cloud at another 
wavelength by using the equation for $\tau$ above.
We altered COLTEMP to allow  
temperatures as low as 10K to be used (the publicly available version 
only accepts temperatures greater than 30K).  
In this way, optical depth maps were made of our chosen regions
at 100~$\mu$m.

\begin{figure*}
 \centering
  \begin{minipage}{160mm}
\epsfxsize=16cm
\epsfbox[80 170 880 630]{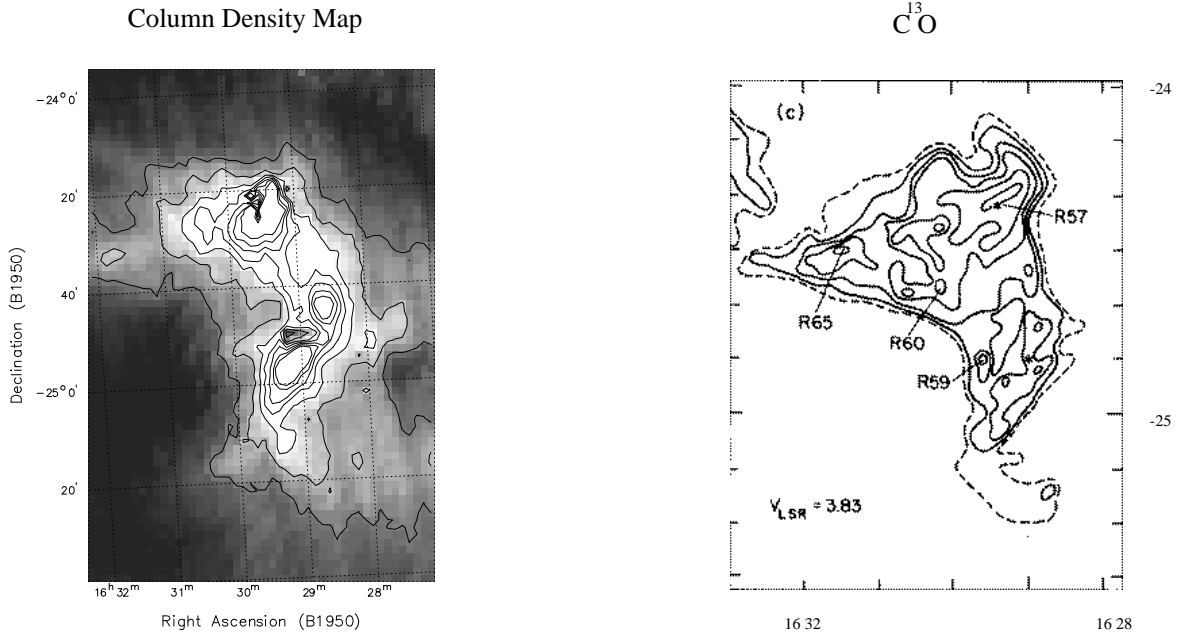}
\caption{(a) 100-$\mu$m optical depth map of the molecular cloud L1689,
produced from the ISSA sky survey plates.
Base contour level is 0.8 $\times 10^{-3}$ and the contour interval is
0.3 $\times 10^{-3}$. (b) $ \rm ^{13}$CO map of the same field taken 
from Loren (1989).}
\end{minipage} 
\end{figure*}

\subsection{Optical depth maps}

The optical depth at 100$ \rm \mu$m due to dust along the line of sight is 
expressible (Hildebrand 1983) as:

\[ \tau_{100}=\pi \langle a \rangle^2 Q_{100}N_{\rm g}, \]

\noindent
where $ N_{\rm g}$ is the column density of grains, $ \langle a \rangle$ is
the average radius of the 
grains and $ Q_{100}$ is the emission efficiency of the dust grains 
at 100$ \rm \mu$m. The mass column density of the grains is then given by:

\[ \sigma_{\rm d}=\frac{4}{3}\left(\frac{a\rho}{Q_{100}}\right)\tau_{100},\]

\noindent
where $ \rho$ is the average grain density. The total mass of 
the dust in a cloud of projected area A is therefore given by:

\[ M_{\rm{dust}}=A\langle \sigma_d \rangle , \]

\noindent
where $\langle \sigma_d \rangle$ is the 
mean column density of dust in the cloud. 
The typical dust-to-gas mass ratio in the local ISM is
normally assumed to be approximately 
1:100. However it is clear that at 100~$\mu$m 
a significant fraction of the cold dust in the ISM is not detected,
due to temperature and optical depth effects (see e.g. Wood et al. 
1994). Hence a simple application of this ratio to these data will
underestimate the total mass of gas.
Wood et al. (1994) argued that
only 1/50 of the total dust is detected 
at {\it IRAS} wavelengths, and used
a dust-to-gas ratio of 1:2000 in their analysis.
They arrived at this value
by using two relations: Firstly they adopted 
$A_{\rm v} \sim 2\times 10^4 \times \tau_{100}$ (Langer et al. 1989); 
and secondly they used
$  N_{\rm H_2}(cm^{-2})\sim 10^{21} \times A_{\rm V}$ 
(Bohlin, Savage \& Drake 1978).
We follow Wood et al. (1994) and use a
value of 1/2000 in the current study. 
With a value for 
$ (a\rho/Q_{100}) \sim 32$gcm$^{-1}$, one obtains the expression for the
mass of material in a cloud:

\[ M_{cloud}/{\rm M_\odot} \sim 
1.25\times 10^{-2} \times D^2(\rm{pc}) \sum_{\rm{cloud}}\tau_{100}, \]

\noindent
where $D$ is the distance to the cloud, 
and the optical depth of each pixel in the cloud image is summed.

\subsection{Cloud selection}

Of the resultant optical depth maps,
some contained from one to three cloud complexes,
varying in size from a few pixels to half a field,
while several more
were discarded, either due to having very little structure, 
being too noisy, or having major
contamination from residual
Zodiacal emission that none of the processing had been able to remove.
Some of the clouds were found at the edges of the fields and hence 
were truncated. 
>From the remaining maps, 17 clouds were
selected for further study.
The clouds we selected are listed in Table 1: Column 1 
lists the name we assigned to each cloud,
derived from the ISSA field number in which 
it was found; Columns 2 \& 3 give the approximate position of the
centre of the cloud; and the remainder of the table assigns distances,
velocities and associations to the clouds we have selected.

It was found that 3 of the clouds were previously identified by Lynds as 
dark clouds (Lynds 1962), 2 contained Lynds bright nebulae (Lynds 1965), 
7 had been identified by other authors (Taylor et al. 1987; 
Ramesh 1994), who had subsequently measured their CO velocities, and 3 
had nearby open cluster associations. Of the 3 open clusters, two have been 
dated and were found to be old: NGC 7142 is
thought to be 4 billion years old; 
Merlotte 66 is 6 billion years old; in both cases
implying that they were probably not linked with the cloud. In addition,
comparison with the CO Galactic plane surveys (Dame et al. 1987) revealed 
that several clouds were associated 
with known cloud complexes. Only two had no previously published 
associations.
Column 4 of Table 1 lists the known cloud associations.

Unlike molecular maps and surveys, which give the velocity of the clouds 
and hence give an estimate of the distances, these ISSA selected clouds
-- like the optically selected clouds of Lynds (1962) -- do not
have easily derivable distances. 
Distances were estimated either from velocity and spatial association 
with the Orion, Cepheus, Chameleon and Ophiuchus 
complexes, or by estimating an
upper limit for distance obtained by assuming the clouds 
lie in the Galactic
Disc -- i.e. less than 60 pc away from the Galactic Plane (c.f. Clemens, 
Sanders \& Scoville 1988). 
Clouds found in Cepheus present a particular problem when one attempts to 
assign a distance. There are 
two different complexes along the line of sight: one at 
approximately 300~pc 
with a velocity $ \rm \sim$ 0~kms$^{-1}$;
and one in the local spiral arm at approximately 800~pc and with a 
velocity of $ \rm \sim$ $-$12kms$^{-1}$ (Grenier et al. 1989). 
Some of the Cepheus clouds we selected had  
been sampled with CO observations (Taylor et al. 1987), 
and the measured velocities revealed the clouds belonged to one or other
complex. This revealed that the sample presented here contains clouds from 
both complexes and hence that the three Cepheus
clouds in the sample without CO observations could
be at either of the two distances.
In total, 11 of the 16 clouds had a 
single distance assigned to them, 2 had 
upper limits, and 3 could be at either of 2 distances.
Column 5 of Table 1 gives the velocity 
of the cloud (if previously measured), 
and column 6 gives the estimated distance. 

\subsection{The case of L1689}

As a cross-check of both
the procedure used and the mass estimates derived,
we made a map of a previously studied cloud, L1689. 
Colour temperature and optical depth maps were constructed 
and the results are presented in Figure~2(a) as a 100-$\mu$m optical
depth map of the region. An extended region of high 100-$\mu$m optical
depth containing some structure is seen. 
Figure~2(b) shows a $^{13}$CO map of the same region taken from Loren (1989).
A similar morphology is seen in both maps.
The two star-forming cores R57 (alias L1689N), and R59 (alias L1689S) are 
clearly visible, as is the isolated core R65 (alias L1689B).

At the centre of the cloud there is an apparent 
`hole' in the optical depth map at the position of the bright
point source, IRAS16288, which is a young 
protostar at the centre of L1689S. 
Wood et al. (1994) also noted that
a bright, point-like source dominating
the {\it IRAS} emission can cause an apparent hole in the
100-$\mu$m optical depth. 
They ascribed this to a beam filling factor effect.
The right hand side of the above equation for $\tau_{(100)}$ also contains
a term for the solid angle of the source, $\Omega$,
which has been set equal to 1 in our analysis. For an extended
molecular cloud source this is a valid assumption, but it fails when
there is a bright point source in the beam. This is the effect we see
in the case of IRAS16288.

We calculated the mass of L1689, 
using the above equation, from the 100-$\mu$m optical depth map
in Figure 2(a), and found a value of 448~M$_\odot$
(assuming a distance of 160 pc).
Loren (1989) used the $\rm ^{13}$CO data shown in Figure 2(b), and
obtained a value of 566~M$_\odot$.
These two measurements are consistent
to within $\pm$20 per cent, 
which is as accurate as either method can claim,
and so we conclude that the technique outlined not only gives accurate 
qualitative information on the morphology of interstellar clouds, but 
also appears to provide
relatively good quantitative estimates of cloud masses.
Nonetheless, we realise that masses derived from 100-$\mu$m optical
depths might be underestimated in some cases, if the 100-$\mu$m
emission is completely optically thick. In none of the cloud cores
that we studied did this appear to be the case.

\section{The core catalogue}

\subsection{Core properties}

Figures 3 \& 4 show grey-scale images, with isophotal contours overlaid,
of nine of our molecular cloud regions.  
They can be seen to vary in size from 1 to 5 degrees across, and also to vary
in complexity and structure. Some residual striping can be seen in two of the
fields. However the typical structures associated with molecular clouds can be
seen in all of the images -- namely cores, filaments and other structures on
all scales within the maps. 
Some of the structure seen in the maps corresponds to previously known
molecular clouds, but some of the clouds, and many of the cores, had not been
previously catalogued. For example, Figure 3(d), 420B, and Figure 4(d), 423A,
both contain a Lynds dark nebula and an open cluster. 423A also contains 
an HII region and two further catalogued clouds. Some of the regions contain
no previous associations -- especially those in the southern hemisphere. 
For a full list of known
associations see Table 1 and section 2.4 above. 

 \begin{figure*}
\begin{tabular}{ccccccc}
\multicolumn{4}{c}{\raisebox{7cm}[0cm][0cm]{(a) \epsfxsize=9cm
\epsfbox[26 200 470 540]{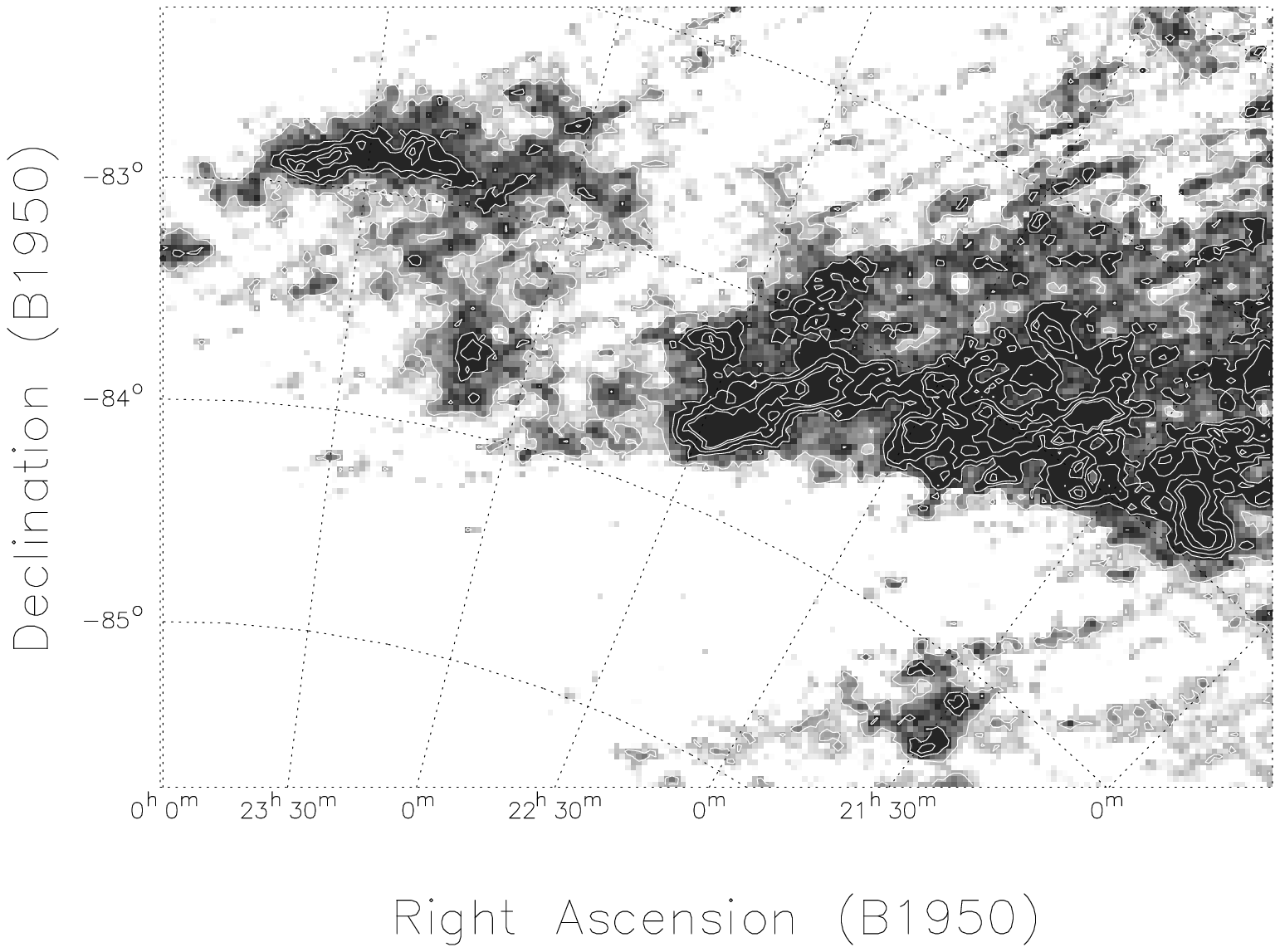}} } &
\multicolumn{3}{c}{(b) \epsfxsize=7cm
\epsfbox[90 60 390 670]{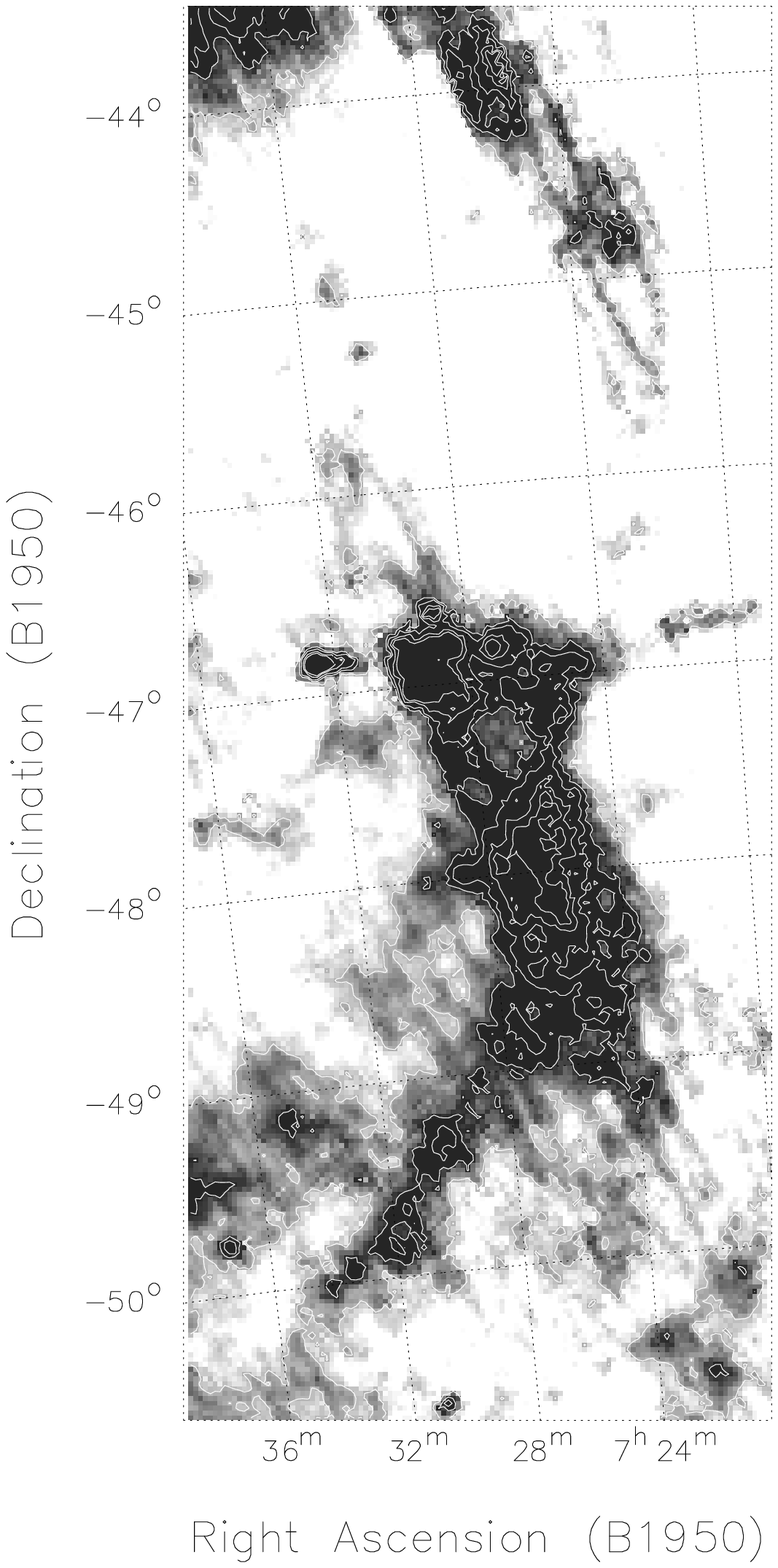} }\\
\multicolumn{4}{c}{\raisebox{8cm}[0cm][0cm]{(c) \epsfxsize=9cm
\epsfbox[40 180 460 550]{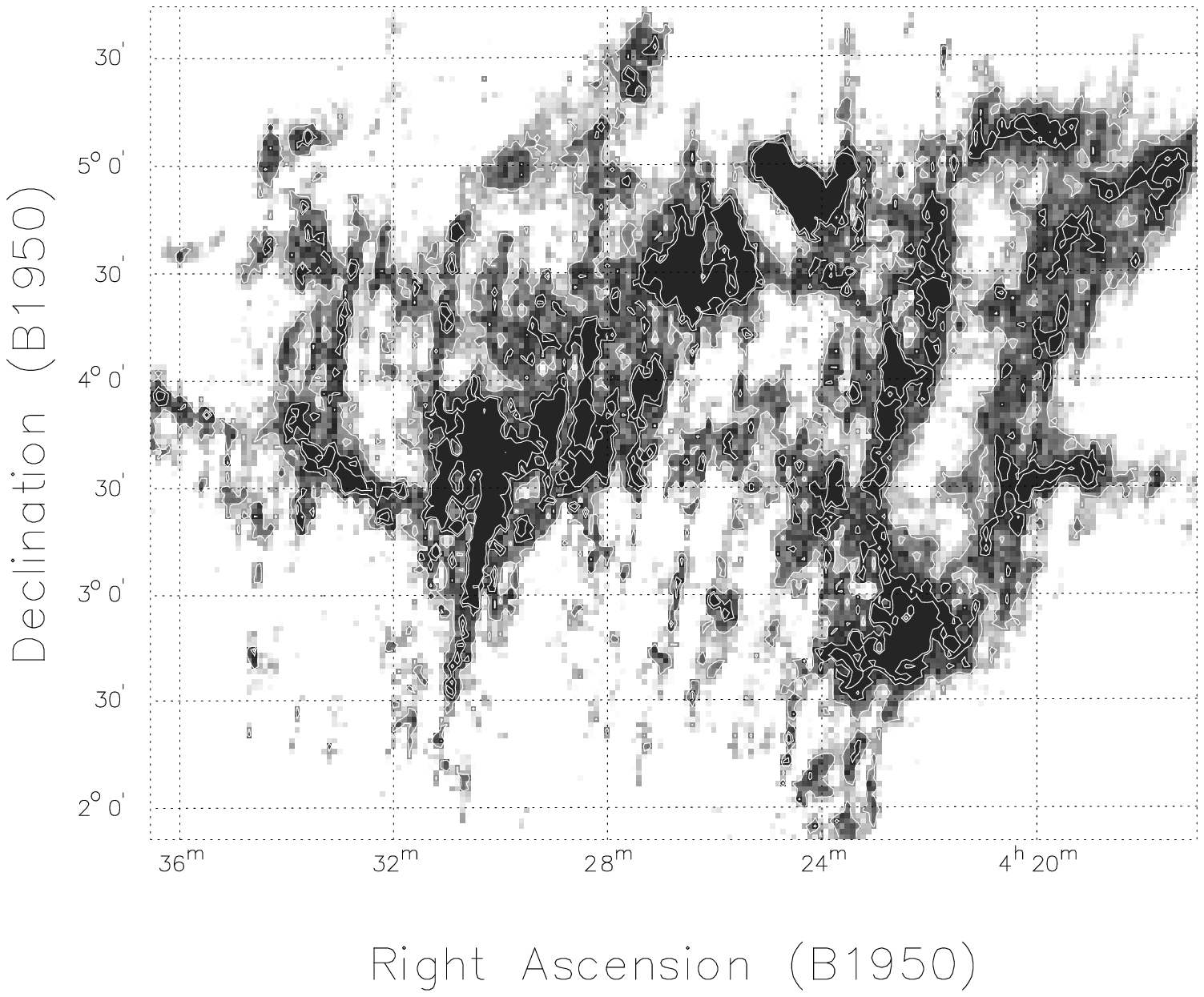}} } &
\multicolumn{3}{c}{(e) \epsfxsize=7cm
\epsfbox[80 170 400 570]{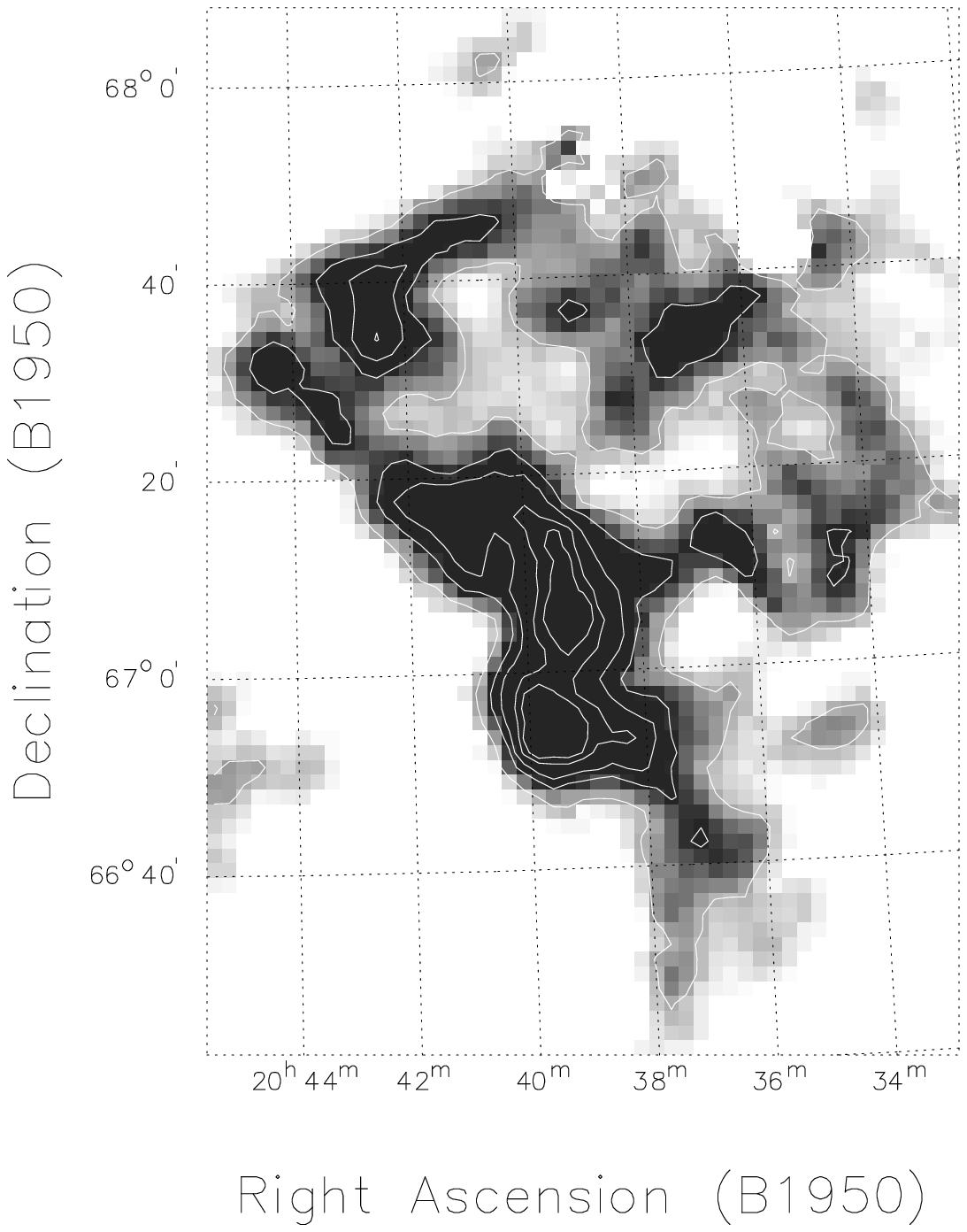} }\\
\multicolumn{4}{c}{\raisebox{2cm}[0cm][0cm]{(d) \epsfxsize=9cm
\epsfbox[10 200 480 530]{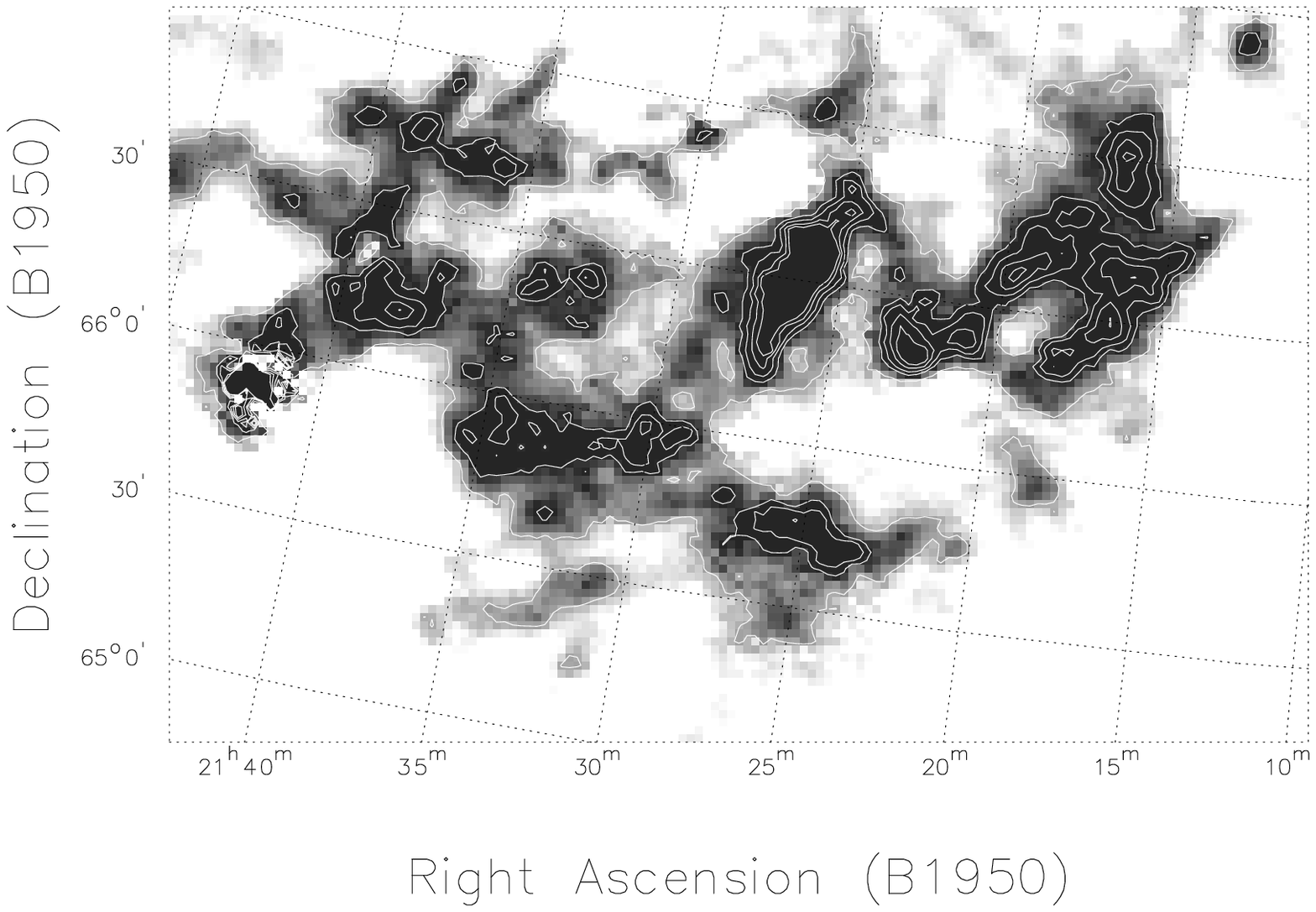}} } &
\multicolumn{3}{c}{} \\
\end{tabular}

\caption[Column density maps of some clouds]{100-$\mu$m optical
depth maps, with contours
overlaid, of: (a) 002A; (b) 050A; (c) 205A; (d) 420B; (e) 420C.}
\end{figure*}

\begin{figure*}
\begin{tabular}{cccc}
\multicolumn{2}{c}{(a) \epsfxsize=8cm
\epsfbox[80 50 420 690]{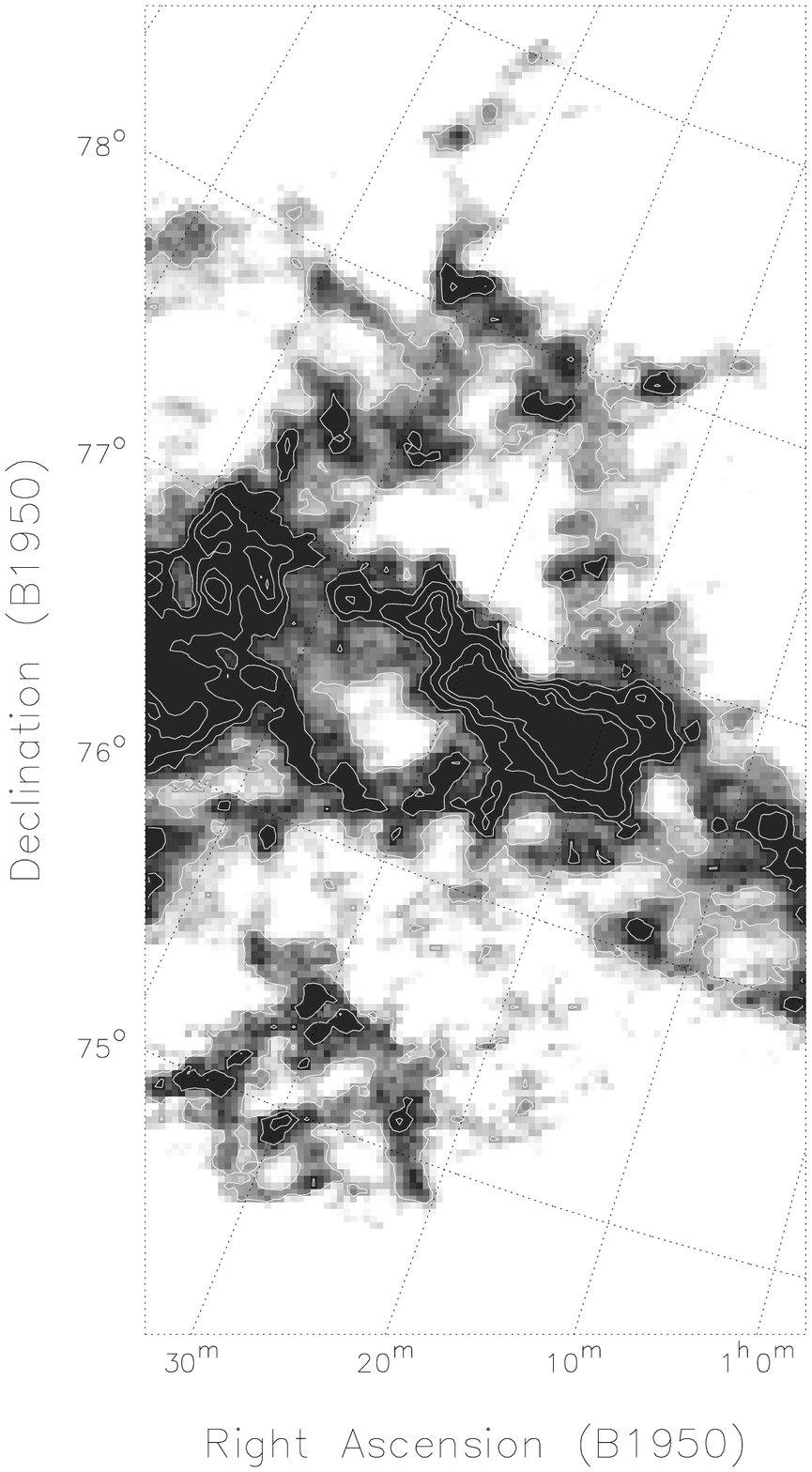} } &
\multicolumn{2}{c}{\raisebox{1cm}[0cm][0cm]{(b) \epsfxsize=8cm
\epsfbox[50 80 410 660]{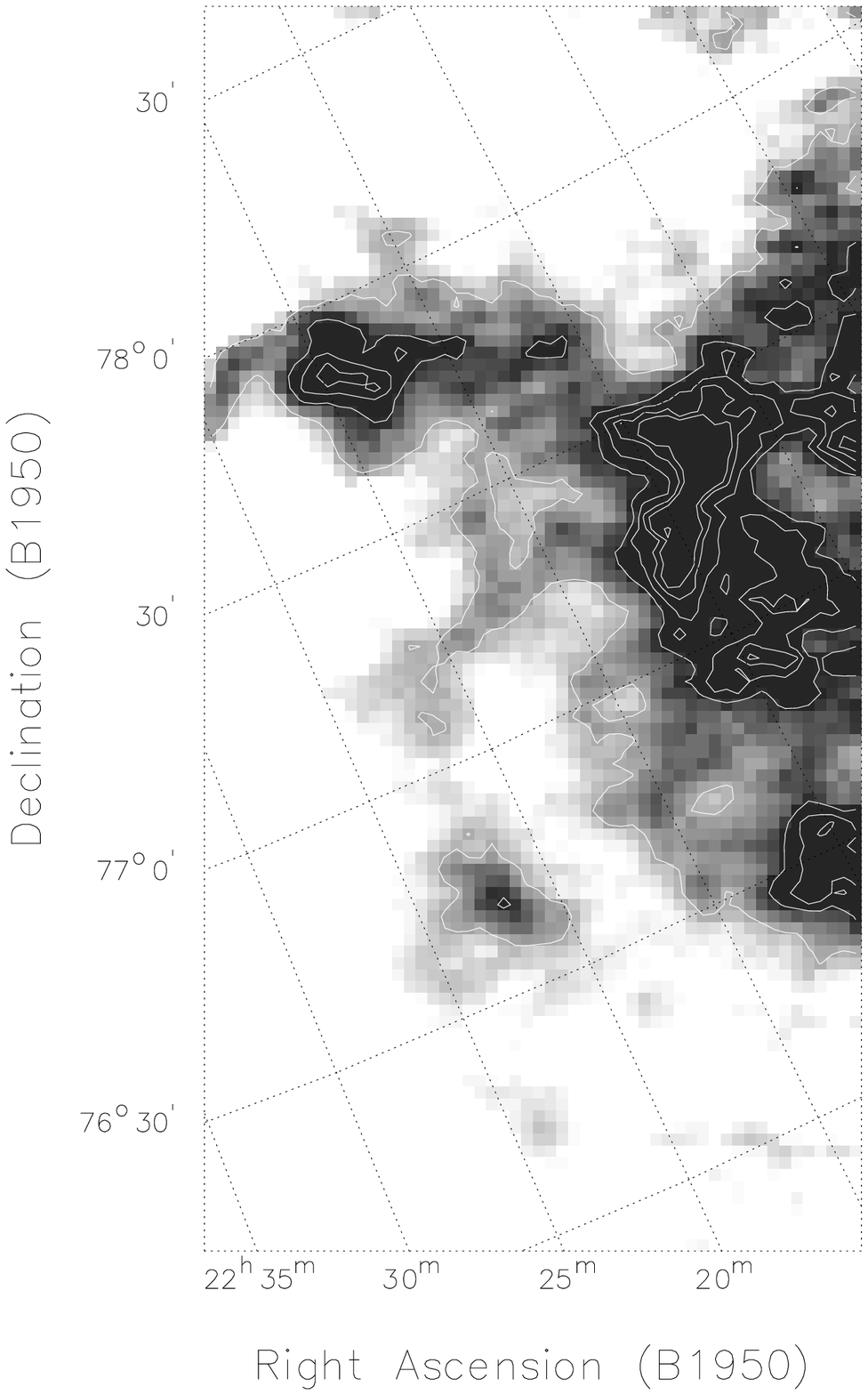}} }\\
\multicolumn{2}{c}{\raisebox{1cm}[0cm][0cm]{(c) \epsfxsize=8cm
\epsfbox[50 230 430 520]{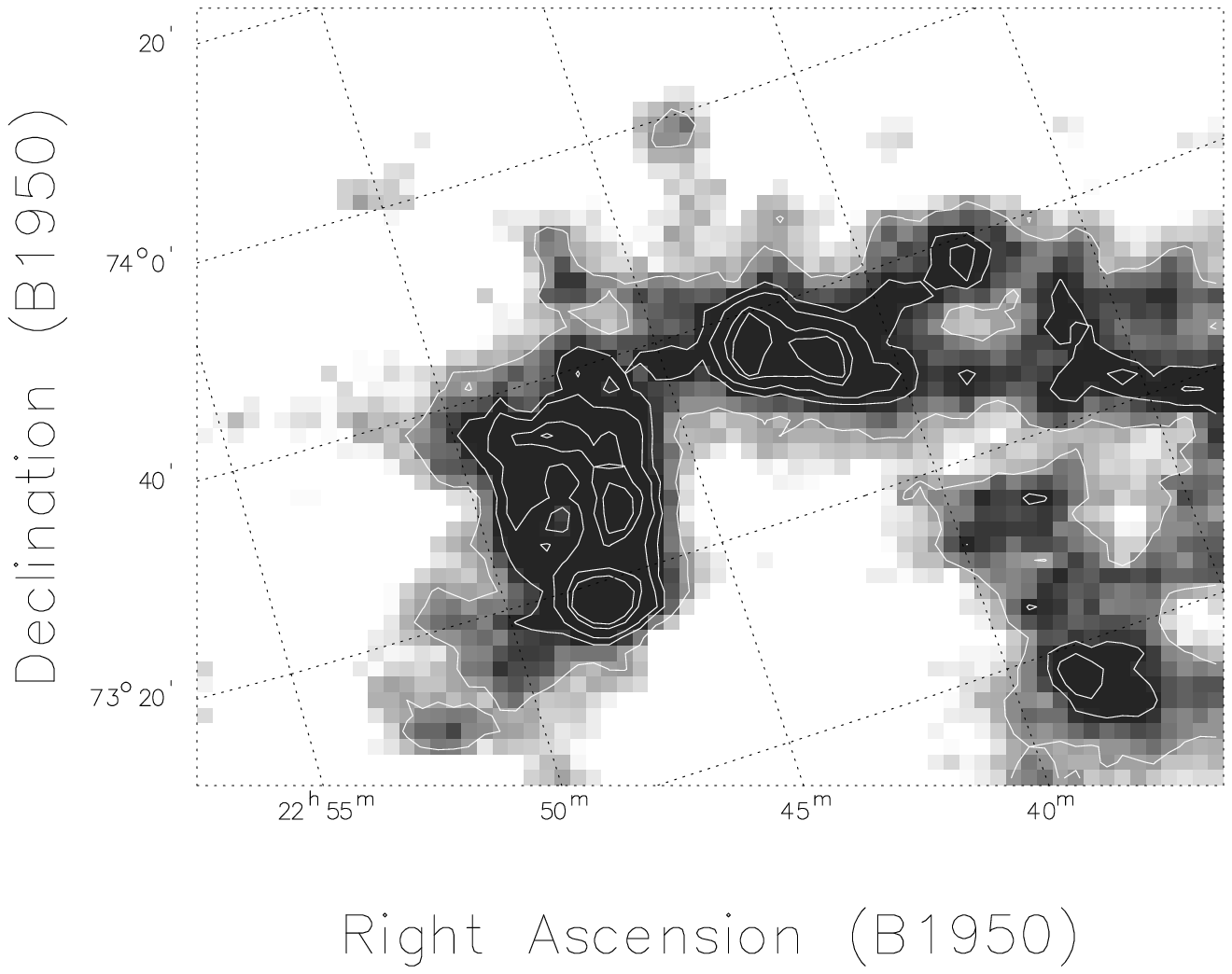}} } &
\multicolumn{2}{c}{(d) \epsfxsize=8cm
\epsfbox[60 180 440 560]{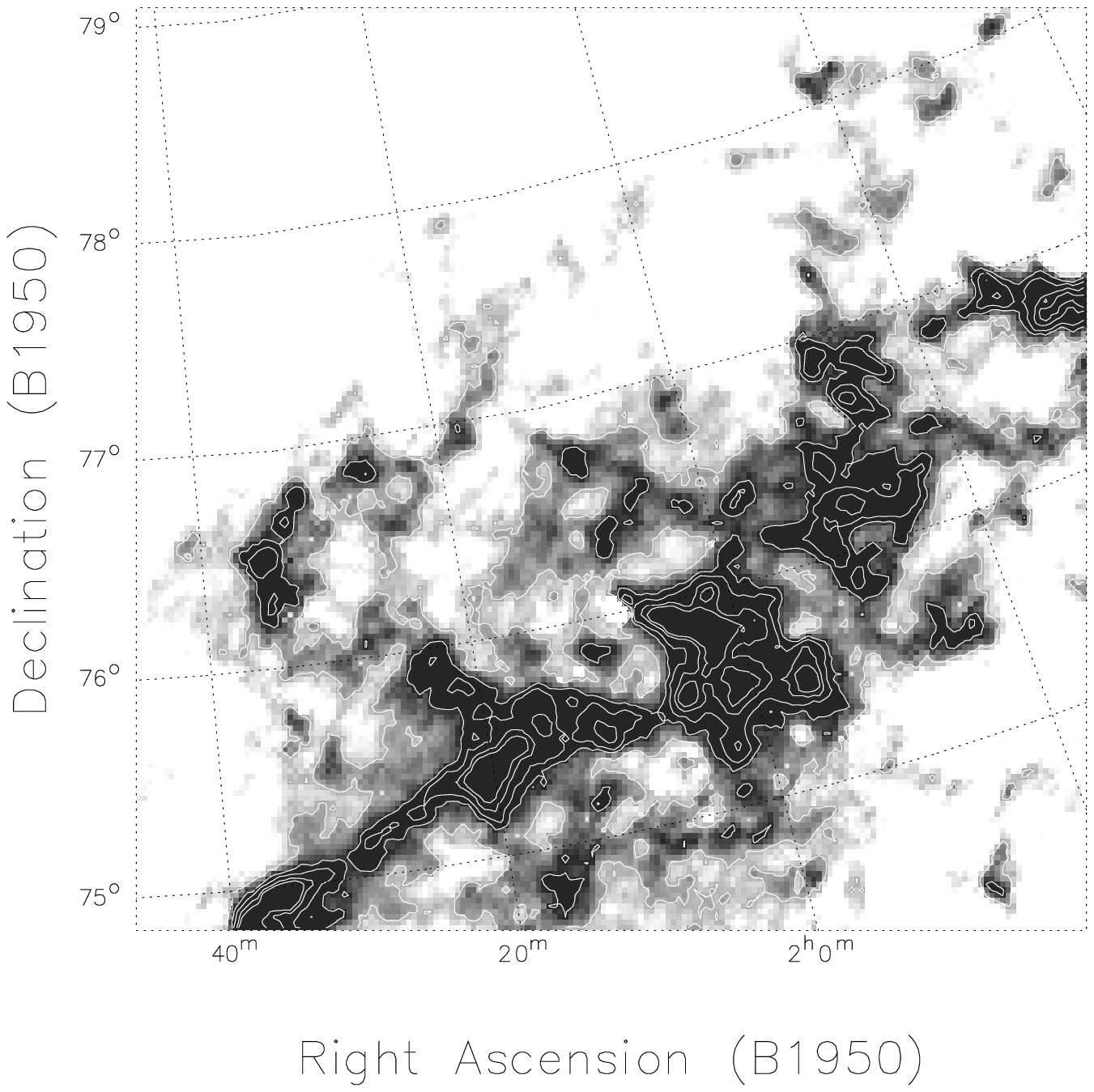} }\\
\end{tabular}

\caption[Column density maps of some clouds]{100-$\mu$m optical depth
maps, with contours
overlaid, of: (a) 422A; (b) 422C; (c) 422B; (d) 423A}
\end{figure*}

A catalogue of the most opaque regions in each cloud was produced and
the maps were examined. The definition of what constitutes a core in a
molecular cloud is of necessity somewhat subjective.
We chose to define a core as 
the most opaque 1\% of a cloud's area.
This level was chosen to ensure that the area defined as a core was
relatively small compared to the cloud, and hence would only include the
most dense regions where we would expect star formation to take place. 
We then calculated the mass of each core.

\begin{center}
\begin{table*}
%\begin{minipage}{220mm}
\caption{The catalogue of molecular cloud cores, together with
their positions, sizes, optical depths, shapes, orientations and masses.}
\begin{tabular}{cccccccccccc}
\hline
JWT & Cloud & R.A. & Dec. & Temp & Size & 
$ \rm \overline{\tau} $ &  $ \rm \tau_{peak}$ & 
Ellip- & $ \overline{\rm r}$ & $\theta$ & Mass  \\
No. & name & (1950) & (1950) & (K) & (arcmin$^2$) & 
($\times 10^{-3}$) & ($\times 10^{-3}$) & ticity &
(pc) & ($^\circ$) & (M$_\odot$)  \\ \hline
1 & 270A  &   00$^{\rm h}$ 03$^{\rm m}$ 03$^{\rm s}$ &
 $+$13$^{\circ}$ 56$^{\prime}$ 30$^{\prime\prime}$ & 29.2 & 
23.4 & 0.02 &  0.018 &    0.43  & $<$0.2 &   11 & $<$2    \\ 
2 & 422A  &   01$^{\rm h}$ 21$^{\rm m}$ 38$^{\rm s}$ &
   $+$76$^{\circ}$ 41$^{\prime}$ 37$^{\prime\prime}$ &  21.7 & 
105 &  0.40 &   0.439 &  0.43  &  0.5--1.2 &  67 & 300--2100  \\ 
3 & 422A  &   01$^{\rm h}$ 27$^{\rm m}$ 54$^{\rm s}$ &
   $+$76$^{\circ}$ 44$^{\prime}$ 27$^{\prime\prime}$ &  21.5 & 
45.3 & 0.38 &  0.404 & 0.71  & 0.3--0.9 &  68 & 120--880    \\ 
4 & 423A  &   02$^{\rm h}$ 03$^{\rm m}$ 18$^{\rm s}$ &
    $+$75$^{\circ}$ 51$^{\prime}$ 50$^{\prime\prime}$ &   20.8 & 
54.7 & 0.52 &    0.580 &    0.68  & 0.38 &   71 & 200    \\ 
5 & 423A  &   02$^{\rm h}$ 20$^{\rm m}$ 46$^{\rm s}$ &
    $+$75$^{\circ}$ 22$^{\prime}$ 54$^{\prime\prime}$ &  21.4 & 
12.5 & 0.49  &  0.505 &    0.18  & 0.18 &  134 &  44   \\ 
6 & 002B  &   02$^{\rm h}$ 31$^{\rm m}$ 55$^{\rm s}$ &
  $-$84$^{\circ}$ 56$^{\prime}$ 03$^{\prime\prime}$ & 23.8 & 
21.9 & 0.12 &    0.142 &    0.01  & 0.16 &  123 & 8.3    \\ 
7 & 423A  &   02$^{\rm h}$ 36$^{\rm m}$ 20$^{\rm s}$ &
    $+$74$^{\circ}$ 57$^{\prime}$ 03$^{\prime\prime}$ &  20.2 & 
26.6 &   0.55 &  0.651 &    0.41  & 0.26 &  116  & 100   \\ 
8 & 423B  &   02$^{\rm h}$ 52$^{\rm m}$ 42$^{\rm s}$ &
   $+$81$^{\circ}$ 00$^{\prime}$ 20$^{\prime\prime}$ &   22.0 & 
17.2 & 0.19 &  0.206 &  0.69  &  0.2--0.5 &  35  & 20--160   \\ 
9 & 423B  &   03$^{\rm h}$ 15$^{\rm m}$ 55$^{\rm s}$ &
   $+$81$^{\circ}$ 50$^{\prime}$ 15$^{\prime\prime}$ &   19.8 & 
54.7 &  0.24 & 0.411 &   0.36  & 0.4--0.9 &  120 & 100--690    \\ 
10 & 411A  &   04$^{\rm h}$ 14$^{\rm m}$ 59$^{\rm s}$ &
    $+$64$^{\circ}$ 19$^{\prime}$ 02$^{\prime\prime}$ &  23.0 & 
92.2 & 0.37 &  0.403 &  0.55  & 0.5--1.2 &  167  & 250--1800   \\ 
11 & 411A  &   04$^{\rm h}$ 17$^{\rm m}$ 01$^{\rm s}$ &
     $+$64$^{\circ}$ 07$^{\prime}$ 09$^{\prime\prime}$ &  23.3 & 
12.5 &  0.34 &   0.352 &    0.32  & 0.2--0.5  & 169 & 30--220   \\ 
12 & 411A  &   04$^{\rm h}$ 17$^{\rm m}$ 45$^{\rm s}$ &
    $+$64$^{\circ}$ 14$^{\prime}$ 51$^{\prime\prime}$ &  23.1 & 
29.7 &   0.35 &  0.369 &    0.55  &  0.3--0.7 &  22  & 80--530   \\ 
13 & 411A  &   04$^{\rm h}$ 18$^{\rm m}$ 55$^{\rm s}$ &
   $+$63$^{\circ}$ 53$^{\prime}$ 58$^{\prime\prime}$ &  22.6 & 
79.7 & 0.36 &    0.380 &    0.66  & 0.5--1.1 &  161 & 200--1500 \\
14 & 205A  &   04$^{\rm h}$ 24$^{\rm m}$ 32$^{\rm s}$ &
   $+$04$^{\circ}$ 54$^{\prime}$ 24$^{\prime\prime}$ &   24.1 & 
136 & 0.13 &    0.154 &    0.58  & 0.89 &   38 & 290  \\ 
15 & 205A  &   04$^{\rm h}$ 25$^{\rm m}$ 23$^{\rm s}$ &
   $+$04$^{\circ}$ 30$^{\prime}$ 14$^{\prime\prime}$ &   24.6 & 
21.9 & 0.12 &    0.132 &    0.57  & 0.35 &  173 & 44    \\ 
16 & 205A  &   04$^{\rm h}$ 26$^{\rm m}$ 26$^{\rm s}$ &
   $+$04$^{\circ}$ 30$^{\prime}$ 28$^{\prime\prime}$ &   24.4 & 
57.8 &  0.12 &   0.139 &    0.51  & 0.58 &    3   & 120  \\ 
17 & 205A  &   04$^{\rm h}$ 29$^{\rm m}$ 02$^{\rm s}$ &
   $+$03$^{\circ}$ 48$^{\prime}$ 54$^{\prime\prime}$ &   24.1 & 
21.9 &  0.13 &   0.130 &    0.52  & 0.35 &  168 & 44   \\ 
18 & 205A  &   04$^{\rm h}$ 30$^{\rm m}$ 18$^{\rm s}$ &
    $+$03$^{\circ}$ 37$^{\prime}$ 46$^{\prime\prime}$ &   24.2 & 
26.6 &  0.12 &   0.128 &    0.51  &  0.39 &  71 & 53  \\ 
19 & 205A  &   04$^{\rm h}$ 30$^{\rm m}$ 21$^{\rm s}$ &
   $+$03$^{\circ}$ 18$^{\prime}$ 12$^{\prime\prime}$ &   22.6 & 
31.3 &   0.13 & 0.137 &    0.79  & 0.43 &  168 & 65    \\
20 & 205A  &   04$^{\rm h}$ 30$^{\rm m}$ 32$^{\rm s}$ &
   $+$03$^{\circ}$ 46$^{\prime}$ 53$^{\prime\prime}$ &   24.3 & 
18.8 &   0.12 &  0.130 &    0.51  & 0.32 &  178 & 38   \\ 
21 & 050A  &   07$^{\rm h}$ 26$^{\rm m}$ 05$^{\rm s}$ &
   $-$48$^{\circ}$ 10$^{\prime}$ 15$^{\prime\prime}$ &  26.3 & 
20.3 &  0.18 & 0.188 &  0.55  & $<$0.6 &   39  & $<$200 \\ 
22 & 050A  &   07$^{\rm h}$ 26$^{\rm m}$ 39$^{\rm s}$ &
    $-$47$^{\circ}$ 54$^{\prime}$ 27$^{\prime\prime}$ &  26.0 & 
34.4 &  0.18 & 0.192 &  0.74  & $<$0.8 &   20  & $<$340  \\ 
23 & 050A  &   07$^{\rm h}$ 29$^{\rm m}$ 16$^{\rm s}$ &
   $-$46$^{\circ}$ 54$^{\prime}$ 22$^{\prime\prime}$ &  25.9 & 
170 & 0.20 & 0.278 & 0.28  & $<$1.9 &  148  & $<$1900  \\ 
24 & 050A  &   07$^{\rm h}$ 32$^{\rm m}$ 30$^{\rm s}$ &
   $-$46$^{\circ}$ 50$^{\prime}$ 12$^{\prime\prime}$ &  25.5 & 
26.6 &  0.20 & 0.240 &  0.30  & $<$0.7 &   97 & $<$300   \\ 
25 & 221A  &   15$^{\rm h}$ 30$^{\rm m}$ 12$^{\rm s}$ &
  $-$03$^{\circ}$ 23$^{\prime}$ 34$^{\prime\prime}$ &   25.0 & 
40.6 & 0.09 &    0.101 &    0.31  & 1.35 &   50 & 4.3  \\ 
26 & 221A  &   15$^{\rm h}$ 39$^{\rm m}$ 03$^{\rm s}$ &
  $-$03$^{\circ}$ 24$^{\prime}$ 15$^{\prime\prime}$ &  25.1 & 
43.8 &  0.07 & 0.079 &    0.46  & 0.14 &   19 & 4.1   \\ 
27 & 221A  &   15$^{\rm h}$ 40$^{\rm m}$ 36$^{\rm s}$ &
   $-$03$^{\circ}$ 16$^{\prime}$ 05$^{\prime\prime}$ &  25.0 & 
18.8 &  0.08 &    0.077 &    0.70  & 0.09 &    4   & 1.8 \\
28 & 221A  &   15$^{\rm h}$ 43$^{\rm m}$ 25$^{\rm s}$ &
  $-$03$^{\circ}$ 50$^{\prime}$ 38$^{\prime\prime}$ & 25.4 & 
28.1 & 0.08 &     0.081 &    0.60  &  0.11 &  14 & 2.6    \\ 
29 & 221A  &   15$^{\rm h}$ 44$^{\rm m}$ 21$^{\rm s}$ &
  $-$03$^{\circ}$ 49$^{\prime}$ 22$^{\prime\prime}$ & 05.6 & 
50.0 &  0.08 &    0.086 &    0.69  & 0.15 &   19 & 4.8  \\ 
30 & 021A  &   18$^{\rm h}$ 03$^{\rm m}$ 10$^{\rm s}$ &
   $-$75$^{\circ}$ 28$^{\prime}$ 48$^{\prime\prime}$ & 25.6 & 
259 & 0.06 &     0.070 &    0.60  & 0.55 &  169  & 50 \\ 
31 & 021A  &   18$^{\rm h}$ 13$^{\rm m}$ 05$^{\rm s}$ &
   $-$75$^{\circ}$ 28$^{\prime}$ 21$^{\prime\prime}$ &  25.9 & 
46.9 & 0.06 &     0.063 &    0.63  & 0.23 &  174  & 8.7 \\ 
32 & 021A  &   18$^{\rm h}$ 33$^{\rm m}$ 20$^{\rm s}$ &
   $-$76$^{\circ}$ 00$^{\prime}$ 14$^{\prime\prime}$ & 25.6 & 
46.9 &   0.06 &   0.063 &    0.11  & 0.23 &  141 & 8.7  \\ 
33 & 021A  &   18$^{\rm h}$ 33$^{\rm m}$ 22$^{\rm s}$ &
  $-$74$^{\circ}$ 16$^{\prime}$ 04$^{\prime\prime}$ &  25.5 & 
37.5 &  0.06 &    0.061 &    0.36  & 0.21 &   11 & 6.8  \\ 
34 & 021A  &   18$^{\rm h}$ 33$^{\rm m}$ 48$^{\rm s}$ &
  $-$75$^{\circ}$ 45$^{\prime}$ 39$^{\prime\prime}$ &  25.5 & 
20.3 & 0.06 &     0.063 &    0.61  & 0.15 &  173 & 3.8 \\ 
35 & 021A  &   18$^{\rm h}$ 37$^{\rm m}$ 53$^{\rm s}$ &
  $-$74$^{\circ}$ 00$^{\prime}$ 43$^{\prime\prime}$ &  25.0 & 
12.5 & 0.06  &   0.056 &    0.66  & 0.12 &  157 & 2.2  \\
36 & 420C  &   20$^{\rm h}$ 39$^{\rm m}$ 20$^{\rm s}$ &
  $+$67$^{\circ}$ 07$^{\prime}$ 58$^{\prime\prime}$ &   19.1 & 
40.6 & 0.60  &  0.680 &    0.62  & 0.32 &  175  & 160   \\ 
37 & 420C  &   20$^{\rm h}$ 39$^{\rm m}$ 37$^{\rm s}$ &
  $+$66$^{\circ}$ 54$^{\prime}$ 51$^{\prime\prime}$ &   19.0 & 
35.9 & 0.64 &    0.761 &    0.25  & 0.31 &   39  & 170   \\
38 & 420B  &   21$^{\rm h}$ 16$^{\rm m}$ 11$^{\rm s}$ &
  $+$66$^{\circ}$ 39$^{\prime}$ 03$^{\prime\prime}$ &   23.6 & 
29.7 & 0.35 &   0.359 &    0.44  & 0.35 &   74 & 530    \\ 
39 & 420B  &   21$^{\rm h}$ 16$^{\rm m}$ 42$^{\rm s}$ &
  $+$66$^{\circ}$ 29$^{\prime}$ 40$^{\prime\prime}$ &   23.9 & 
20.3 & 0.35 &    0.359 &    0.20  & 0.62 &  130  & 360    \\
40 & 420B  &   21$^{\rm h}$ 16$^{\rm m}$ 43$^{\rm s}$ &
  $+$66$^{\circ}$ 59$^{\prime}$ 26$^{\prime\prime}$ &   23.3 & 
25.0 & 0.35 &   0.369 &    0.59  & 0.67 &  172 & 450    \\   
41 & 002A  &   21$^{\rm h}$ 18$^{\rm m}$ 29$^{\rm s}$ &
  $-$82$^{\circ}$ 49$^{\prime}$ 34$^{\prime\prime}$ &  22.7 & 
59.4 & 0.15 &    0.164 &    0.66  & 0.26 &   16 & 29  \\ 
42 & 420B  &   21$^{\rm h}$ 19$^{\rm m}$ 34$^{\rm s}$ &
  $+$66$^{\circ}$ 37$^{\prime}$ 23$^{\prime\prime}$ &   23.3 & 
18.8 & 0.35 &    0.371 &    0.62  & 0.58 &  120 & 340   \\
43 & 420B  &   21$^{\rm h}$ 22$^{\rm m}$ 38$^{\rm s}$ &
  $+$66$^{\circ}$ 21$^{\prime}$ 22$^{\prime\prime}$ &   24.5 & 
37.5 & 0.36 &    0.393 &    0.38  & 0.84 &   34 & 690    \\ 
44 & 420A  &   21$^{\rm h}$ 26$^{\rm m}$ 21$^{\rm s}$ &
  $+$66$^{\circ}$ 30$^{\prime}$ 46$^{\prime\prime}$ &  24.0 & 
177 & 0.39 &    0.514 &    0.65  & 1.80 &  149 & 3600     \\ 
45 & 002A  &   21$^{\rm h}$ 38$^{\rm m}$ 55$^{\rm s}$ &
   $-$82$^{\circ}$ 45$^{\prime}$ 12$^{\prime\prime}$ &  22.4 & 
54.7 &  0.16 &  0.186 &    0.47  & 0.25 &  100 & 28    \\
46 & 420B  &   21$^{\rm h}$ 42$^{\rm m}$ 06$^{\rm s}$ &
  $+$65$^{\circ}$ 53$^{\prime}$ 12$^{\prime\prime}$ &  43.6 & 
34.4 & 0.50 &   1.030 &    0.34  &  0.79 &  62 & 880    \\ 
47 & 334A  &   21$^{\rm h}$ 46$^{\rm m}$ 35$^{\rm s}$ &
   $+$35$^{\circ}$ 20$^{\prime}$ 01$^{\prime\prime}$ & 27.5 & 
15.6 &  0.06   & 0.063 &    0.39  & $<$0.6 &  145  & $<$60   \\ 
48 & 334A  &   21$^{\rm h}$ 46$^{\rm m}$ 51$^{\rm s}$ &
  $+$35$^{\circ}$ 00$^{\prime}$ 58$^{\prime\prime}$ & 27.2 & 
48.4 & 0.06 & 0.071 &    0.22  & $<$1.0 & 118  & $<$200    \\ 
49 & 002A  &   21$^{\rm h}$ 47$^{\rm m}$ 00$^{\rm s}$ &
  $-$83$^{\circ}$ 00$^{\prime}$ 13$^{\prime\prime}$ &  22.5 & 
42.2 & 0.15 &  0.172 &    0.62  & 0.22 &  95  & 21 \\  
50 & 334A  &   21$^{\rm h}$ 47$^{\rm m}$ 31$^{\rm s}$ &
   $+$35$^{\circ}$ 23$^{\prime}$ 19$^{\prime\prime}$ & 27.3 & 
50.0 &  0.06 &    0.065 &    0.37  & $<$1.1 &   75 & $<$200    \\ 
51 & 334A  &   21$^{\rm h}$ 48$^{\rm m}$ 55$^{\rm s}$ &
  $+$35$^{\circ}$ 46$^{\prime}$ 40$^{\prime\prime}$ & 27.5 & 
15.6 &  0.06 &    0.067 &    0.35  & $<$0.6 &  153 & $<$60    \\
52 & 002A  &   21$^{\rm h}$ 58$^{\rm m}$ 57$^{\rm s}$ &
  $-$83$^{\circ}$ 14$^{\prime}$ 17$^{\prime\prime}$ &  23.0 & 
28.1 & 0.15 &    0.160 &    0.23  & 0.18 &   83 & 14   \\ 
53 & 002A  &   21$^{\rm h}$ 59$^{\rm m}$ 28$^{\rm s}$ &
  $-$83$^{\circ}$ 00$^{\prime}$ 02$^{\prime\prime}$ &  22.4 & 
21.9 &   0.15 &  0.156 &    0.41  &  0.16 &  72 & 11    \\ 
54 & 422C  &   22$^{\rm h}$ 07$^{\rm m}$ 50$^{\rm s}$ &
 $+$77$^{\circ}$ 22$^{\prime}$ 11$^{\prime\prime}$ &   21.4 & 
25.0 & 0.30 &    0.320 &    0.48  & 0.25 &  180 & 55    \\ 
55 & 422C  &   22$^{\rm h}$ 09$^{\rm m}$ 58$^{\rm s}$ &
   $+$77$^{\circ}$ 15$^{\prime}$ 03$^{\prime\prime}$ &   21.8 & 
12.5 &  0.30 &   0.321 &    0.61  & 0.18 &  170 & 27   \\ 
56 & 002A  &   22$^{\rm h}$ 15$^{\rm m}$ 50$^{\rm s}$ &
   $-$83$^{\circ}$ 16$^{\prime}$ 45$^{\prime\prime}$ &  23.2 & 
54.7 & 0.15   & 0.170 &    0.48  & 0.25 &  125 & 27    \\ 
57 & 002A  &   22$^{\rm h}$ 21$^{\rm m}$ 53$^{\rm s}$ &
  $-$83$^{\circ}$ 23$^{\prime}$ 50$^{\prime\prime}$ &  23.5 & 
21.9 &   0.15 &  0.163 &    0.57  & 0.16 &  128 & 11    \\
58 & 002A  &   22$^{\rm h}$ 26$^{\rm m}$ 20$^{\rm s}$ &
  $-$83$^{\circ}$ 35$^{\prime}$ 58$^{\prime\prime}$ &  23.0 & 
90.6 & 0.16 &    0.188 &    0.59  & 0.32 &   98  & 48   \\ 
59 & 422B  &   22$^{\rm h}$ 41$^{\rm m}$ 33$^{\rm s}$ &
  $+$73$^{\circ}$ 33$^{\prime}$ 48$^{\prime\prime}$ &  24.1 & 
12.5 & 0.20  &  0.207 &    0.48  & 0.18 &   55 & 18   \\ 
60 & 422B  &   22$^{\rm h}$ 47$^{\rm m}$ 58$^{\rm s}$ &
  $+$73$^{\circ}$ 18$^{\prime}$ 02$^{\prime\prime}$ &  23.8 & 
15.6 &   0.20 &  0.216 &    0.25  & 0.20 &   90 & 23  \\ 
\hline
\end{tabular}
%\end{minipage}
\end{table*}
\end{center}

\begin{center}
\begin{table*}
\caption{{\it IRAS} point sources associated with the cores
in this catalogue, together with their flux densities and other known
associations.}
\begin{tabular}{ccccccc}
\hline
JWT & {\it IRAS} & \multicolumn{4}{c}{Flux densities (Jy)} & Other \\
Number &  Association & 12$ \rm \mu$m & 25$ \rm \mu$m & 
60$ \rm \mu$m & 100 $ \rm \mu$m & associations\\ \hline
2 & 01226+7641 & $ \rm <$0.58 & $ \rm <$0.25 & $ \rm <$0.40 & 2.50 & \\ 
7 & 02368+7453 & 0.44 &  0.89 & 0.89 & 3.21 &  \\ 
8 & 02532+8102& $ \rm <$0.25 & $ \rm <$0.64 & $ \rm <$0.40 & 3.57 &  \\ 
9 & 03139+8151 & 0.33 & $ \rm <$0.25 & $ \rm <$0.40 & $ \rm <$4.99   & \\ 
9 & 03191+8147 & 0.29 & $ \rm <$ 0.25 & $ \rm <$0.40 & $ \rm <$5.23   &   
545K0$\rm ^1$ \\ 
12 & 04178+6416 & 0.67 & $ \rm <$0.25 & $ \rm <$0.40 & $ \rm <$8.18  &  
13118M0$ \rm ^1$ \\ 
14 & 04250+0502 & $ \rm <$0.29 & $ \rm <$0.26 & $ \rm <$0.40 & 2.37  &  \\ 
23 & 07299$-$4644 & $ \rm <$0.25 & $ \rm <$0.25 & 1.05 & 10.34  
& MSH 07408$ \rm ^2$  \\ 
23 & 07300$-$4653 & 0.25 & $ \rm <$0.25 & $ \rm <$1.37 & $ \rm <$10.53 & \\ 
23 & 07305$-$4659 & $ \rm <$0.41 & $ \rm <$0.25 & 0.99 & $ \rm <$10.25  & 
07305$-$4659$ \rm ^3$ \\ 
26 & 15395$-$0330 & $ \rm <$0.33 & $ \rm <$0.85 & $ \rm <$0.40 & 1.02  & \\
34 & 18333$-$7547 & $ \rm <$1.33 & $ \rm <$0.34 & $ \rm <$0.40 & 1.47 &  \\ 
41 & 21185$-$8252 & $ \rm <$0.25 & $ \rm <$0.25 & $ \rm <$0.40 & 2.49 &  \\
43 & 21232+6626 & $ \rm <$2.12 & $ \rm <$0.25 & $ \rm <$0.55 & 9.58 &  \\
46 & 21426+6556 & $ \rm <$0.39 & $ \rm <$0.27 & 2.15 & $ \rm <$21.19 &   \\ 
50 & 21479+3520 & 1.75 & 0.51 & $ \rm <$0.40 & $ \rm <$1.82 & 
DO 20925$ \rm ^4$ \\ 
50 & 21484+3522 & $ \rm <$0.25 & $ \rm <$0.25 & $ \rm <$0.52 & 2.87 &  \\ 
52 & 21580$-$8316& $ \rm <$0.73 & $ \rm <$0.85 & $ \rm <$0.40 & 1.78 &  \\ 
58 & 22278$-$8341& $ \rm <$0.36 & $ \rm <$0.25 & $ \rm <$0.40 & 2.56 & \\ 
60 & 22412+7332 & $ \rm <$0.25 & $ \rm <$0.25 & $ \rm <$0.40 & 2.98 & 
X2241+735$ \rm ^5$ \\ 
60 & 22422+7333& 0.25 & $ \rm <$0.33 & $ \rm <$0.40 & $ \rm <$2.05 &  
RAFGL 2949$ \rm ^6$ \\ \hline
\end{tabular}
\end{table*}
\end{center}

Table 2 lists the properties of our new core catalogue. 
Column 1 lists the new number we have designated for each core in
order of Right Ascension.
Column 2 lists the cloud in which it is located, following the numbering
convention of Table 1.
Columns 3 \& 4 list the position of the centre of each core. 
Column 5 lists the derived colour
temperature of each core.
Column 6 lists the solid angle of each core in square arcmin.
%(as measured by the Starlink \textsc{PISAF} routine), 
Column 7 lists the mean optical depth, and
column 8 lists the peak optical depth of the core.
Column 9 gives the ellipticity, column 10 lists the mean radius, 
column 11 lists the position angle (north through east)
of the major axis and column 12 lists the mass
of each core.

The median 100-$\mu$m optical depth of our 60 cores is 1.9 $\times$
10$^{-4}$, corresponding to a column density of N(H$_2$) = 3.8 $\times$
10$^{21}$ cm$^{-2}$. The median radius of the cores for which a distance
is known is 0.31~pc, with a mean radius of 0.41~pc. The mean is skewed
by a small number of large cores, so we prefer to use the median to
characterise our sample. The median volume density of the sample is
$\sim$2 $\times$ 10$^3$ cm$^{-3}$ (the mean volume density is very
similar). Hence we see that, by selecting our core sample based on a
wavelength of 100~$\mu$m, we have typically selected somewhat
lower density cores than
many previous surveys of molecular cloud cores (see below).
The difference is probably mainly due to our selecting relatively
isolated clouds as a result of our constraints on background emission.

\subsection{PSC associations}

All of the {\it IRAS} point sources within the boundaries of each core 
were located using the Point Source Catalogue (PSC), and
are listed in Table 3. The core in which the source is found is noted 
along with the source flux densities at each {\it IRAS} waveband.
A superscript in the last column
indicates that the {\it IRAS} PSC
listed a known association for the source:
(1) indicates that the source appears in the Smithsonian Astrophysical 
Observatory Star Catalogue (SAO); (2) represents
the catalogue of Ohio State University Radio 
Sources; (3) is the {\it IRAS} Serendipitous Survey Catalogue;
(4) is the Dearborn Observatory Catalog of Faint Red 
Stars; and (5)  is the {\it IRAS} Small Scale
Structure Catalog (Beichman et al. 1988 and references therein).

There are 21 source associations, of which: 11 are detected only at
100~$\mu$m; two are detected only at 60~$\mu$m; with one detected at
60 \& 100~$\mu$m, but no shorter wavelengths;
five are only detected at 12$ \rm \mu$m;
one source is detected at 12 \& 25~$\mu$m, but not at longer wavelengths;
and one source is detected at all four wavebands.
The eleven 100-$\mu$m-only sources have upper limits at the shorter 
wavelengths such that we can say that their spectral energy distributions
peak at a wavelength of around 100~$\mu$m or longer. The one source
detected at all four wavebands and the source detected at 60 \& 100~$\mu$m
only, also have rising spectra to longer
wavelengths. The other two 60-$\mu$m-only sources are also
consistent with having rising spectra.
This is the form of spectral energy density we would
expect for deeply embedded young stellar objects (YSOs) or protostars.
The six sources detected only at the shortest wavelengths have spectra
consistent with field stars or other objects. 
Hence there appear to be 15 PSC sources associated with our
60 cores.

However, some of the PSC associations may still be chance alignments, 
and in addition the 100-$\mu$m-only sources could be cirrus associated
with the clouds rather than embedded YSOs. So
two `control samples' were produced by offsetting the core 
positions by first 2 degrees and then 5 degrees in declination, and 
selecting the {\it IRAS} point sources associated with the new 
positions. In effect this produces two false
populations of cores, one of which is still located in the clouds 
(the 2-degree offset population), and one of which is 
located outside of the clouds. This technique 
was chosen because it was thought that 
it would introduce the least bias in the estimate of the numbers 
of associations that were purely chance alignments.  

In the control sample produced from a 2-degree displacement 
we found ten sources. Four were detected only at 
100$ \rm \mu$m, and
five sources were detected at 12 and 25$ \rm \mu$m -- all of which
were brighter at 12$ \rm \mu$m than 25$ \rm \mu$m. These are thus 
thought to be main sequence field stars -- one had been positively 
identified as a B star. The one remaining source, which was detected 
at all wavelengths, was previously catalogued as a galaxy in the 
Upsalla General Catalogue of Galaxies.
In the sample produced at a displacement of 5 degrees there were 
also ten sources, four of which were 100-$\mu$m-only sources. 
Six of the sources were detected at both 12 and 25~$\mu$m 
and were brighter at 12~$\mu$m than 25~$\mu$m,
and again are probably field stars.

Hence we find no evidence for a significant difference between
the two control samples, either due to the increased displacement, 
or to the location being inside or outside the clouds. Likewise,
we see that the number of foreground stars remains roughly constant
both in the control samples and the real sample. Of the remaining
15 PSC associations in the real sample, we conclude that six
are probably chance alignments.
Of the remainder, up to half may
be 100-$\mu$m-only cirrus sources -- i.e. the
point source might be the cloud core itself (see, for example: 
Benson \& Myers 1989; Reach, Heiles \& Koo 1993;
Bourke et al. 1995b). Hence we estimate that the number of
PSC sources, which are embedded YSOs and are associated 
with our sample of 60 cores, is five.

\section{Comparison with previous core catalogues}

\subsection{Densities}

We can compare our new catalogue of molecular cloud cores with catalogues
produced by previous authors using different methods. For example,
Myers et al. (1983) surveyed 90 cores in C$^{18}$O and $^{13}$CO. Using 
these observations they showed that the cores' C$^{18}$O optical depths, 
as estimated from the ratio of C$^{18}$O brightness to $^{13}$CO
brightness, was reasonably tightly distributed around a mean of 0.35 to 
0.45. They found that the C$^{18}$O column density, $N_{18}$, had a 
mean of $\sim$ 1.6 $\times$ 10$^{15}$ cm$^{-2}$. This led them to estimate
a typical $N(H_2)$ column density of $\sim$ 10$^{22}$ cm$^{-2}$. 

Clemens \& Barvainis (1988) also carried out a survey of molecular
cloud cores, and the {\it IRAS} images of these cores were studied 
by Clemens, Yun \& Heyer (1991). They found a typical 100-$\mu$m optical depth 
for their sample of 2.5 $\times$ 10$^{-4}$. From this we can estimate a 
column density of N(H$_2$) $\sim$
5.0 $\times$ 10$^{21}$ cm$^{-2}$, using the above 
equations. Wood et al. (1994) carried out a survey of the ISSA data for
some known star-forming regions, 
and produced a catalogue of molecular cloud
cores. They found that their cores had a mean column density N(H$_2$)
$\sim$ 4.5 $\times$ 10$^{21}$ cm$^{-2}$.

The mean volume density of the cores in each of these surveys can also 
be estimated. Myers et al. (1983) found a typical volume density in their
cores of $n(H_2) \sim 8 \times 10^3$ cm$^{-3}$. Wood et al. (1994)
did not quote a typical volume density but a value can be estimated from 
their column density if a typical radius for the cores is known. 
The optimum value to 
use is complicated by the fact that they defined cores to be areas with
visual extinction A$_v$ $>$ 4. This leads to the inclusion of several 
very large `cores' which we would define as `clouds'
(their largest `core' is 329 pc$^2$). This skews their mean to a
value much larger than the median. We therefore take the
the second quartile boundary of radius, which is $\sim$ 0.5 pc
in their sample. This leads to an estimate for the typical number 
density of 3 $\times$ 10$^3$ cm$^{-3}$.

The typical number densities of the Clemens \& Barvainis (1988)
cores can also be estimated if the typical size of the cores is known.
Clemens et al. (1991) claim that the typical radius of the cores is 0.35pc. 
This was calculated from the mean solid angle of the cores and an assumed
distance of 600 pc. However, this distance estimate is somewhat uncertain. 
It was derived originally by Clemens \& Barvainis (1988) from two 
considerations: firstly that the cores were generally within 12 degrees 
of the Galactic Plane; and secondly that the cores 
had LSR velocities of between 0 and 10 kms$^{-1}$. 

However, Bourke et al. (1995a)
argued that, because these cores are seen in extinction against the 
background stars of the Galactic Plane, the survey is biased towards 
detecting cores near the Plane. Hence they derive an estimate of 
the typical distance to the cores of 300 pc. Using
this value a typical radius of 0.175pc is derived, and hence a 
volume density of $n(H_2)$ $\sim$ 4.8 $\times$ 10$^3$ cm$^{-3}$ 
is calculated for this sample. This is significantly higher
than the value quoted by Clemens et al. (1991), mainly due to the different
distance assumed, but also partly because we have followed
Wood et al. (1994) in calculating column density from 
100-$\mu$m optical depth. Lemme et al. (1996) studied a subset of the 
Clemens \& Barvainis (1988) cores and reached a similar conclusion:
namely that
Clemens et al. (1991) may have underestimated the typical density
of the cores in the Clemens \& Barvainis (1988) sample.  

Bourke et al. (1995b) presented data for two samples of molecular
cloud cores: one sample consisted of isolated Bok globules, and the
other was a set of cores in more extended clouds. These had mean
column densities of 5 $\times$ 10$^{21}$ cm$^{-2}$ and 1.6 $\times$
10$^{22}$ cm$^{-2}$ respectively, and volume densities of 10$^4$ 
cm$^{-3}$ and 3 $\times$ 10$^4$ cm$^{-3}$ respectively. Hence, it can be
seen that each of these samples has selected cores with somewhat
different properties. We here label the Bourke et al. (1995b) 
extended sample Bourke(1), and the Bok globule sample Bourke(2).
All of the column and volume density estimates have associated
errors from a number of causes. These include chiefly the assumed
fractional abundance of the different tracers used in each set of 
observations. We believe that the values quoted are accurate relative 
to one another to within 20 per cent.
Table 4 summarises the mean properties of each
of the core samples we have discussed.

\begin{center}
\begin{table}
%\begin{minipage}{140mm}
\caption{Summary of mean core properties of previous surveys of 
molecular cloud cores compared to the present work (JWT).
See text for details.}
\begin{tabular}{lcccc}
\hline
Survey & N(H$_2$) & n(H$_2$) & Protostellar & $\tau$ \\ 
 & $\times$10$^{21}$ & $\times$10$^3$ & Percentage &
$\times$10$^6$ \\
 & cm$^{-2}$ & cm$^{-3}$ & & yrs \\
\hline
 Bourke et al.(1) & 16  & 30  & 61$\pm$10  & 0.6$\pm$0.15 \\
 Myers et al.     & 12  & 8   & 44$\pm$7   & 1.3$\pm$0.3  \\
 Bourke et al.(2) & 5.0 & 10  & 36$\pm$7   & 1.7$\pm$0.4  \\
 Clemens et al.   & 5.0 & 5   & 27$\pm$4   & 2.8$\pm$0.6  \\ 
 Wood et al.      & 4.5 & 3   & 23$\pm$3   & 3.3$\pm$0.6  \\
 JWT             & 3.8 & 2   & 9$\pm_{5}^{6}$ & 10$\pm^7_4$  \\ \hline
\end{tabular}
%\end{minipage}
\end{table}
\end{center}

\subsection{Protostellar content and pre-stellar life-time}

A useful parameter in the study of dense cores is the fraction of
cores in a given sample that contain protostars or YSOs. This parameter
can be used to estimate a mean statistical life-time for the sample.
This method was first used by Beichman et al. (1986), who studied 
the embedded YSOs within the core sample of Myers et al. (1983). 
They found that 35 cores had {\it IRAS} sources meeting the 
colour selection criteria of embedded YSOs and 43
had no embedded {\it IRAS} sources. 

This method was also used by Wood et al. (1994) with 
slightly different selection criteria to discard
foreground main sequence stars. 
They found that 59 out of the
255 cores in their sample had at least one embedded source.
We carried out the same test for the
cores in the Clemens \& Baravainis
(1988) sample, and found
65 cores out of 248 have embedded {\it IRAS} sources 
(using the Beichman et al. selection criteria).
Bourke et al. (1995b) found that 27 out of their 76 Bok globules had 
{\it IRAS} point sources satisfying the Beichman criteria, while
36 out of 59 of their cores in extended clouds had embedded sources. 

Beichman et al. (1986) used the percentage of cores with embedded sources
to estimate the lifetime of a core without an embedded YSO by comparing
it with the life-time of the embedded YSO phase. They
found the life-time of a starless core in this way to
be a $\sim \rm 10^6$ years. This was based on assumptions relating to
the time taken for a star to accrete its total mass, and the time
for it to become visible. The uncertainty in this figure is probably
about a factor of 2, but this does not affect the following statistical
arguments, it would simply move the absolute time-scale as a whole
by a factor of 2 (i.e. the absolute values of the vertical axes in
Figures 5 and 6 can move up or down by a factor of 2, but the relative
positions of the data-points do not move -- see below).

Following this method, we here infer a statistical estimate of the 
pre-stellar lifetime, $\tau$, 
of the cores studied in each of the samples 
discussed above. We also make the assumption that in
each survey the cores without embedded sources will go on to form 
protostars in their centres. The cores with embedded sources are 
assumed to have an average lifetime which is 
the same in each sample and equal to the embedded YSO time-scale
derived by Beichman et al. (1986).
This is simply expressible as:

\[ \tau = \frac{\mbox{No. of cores without embedded sources}}
{\mbox{No. of cores with embedded sources}}\times 10^6 \mbox{years}, \]

\noindent
where we have taken the lifetime of cores with embedded sources 
to be $\rm 10^6$ years as discussed above
(c.f. Ward-Thompson et al. 1994).
$\tau$ was calculated for each core sample discussed in
section 4.1, and is listed in Table 4.

\begin{figure}
\begin{center}
\leavevmode
\epsfxsize=8cm
\epsfbox[0 0 500 350]{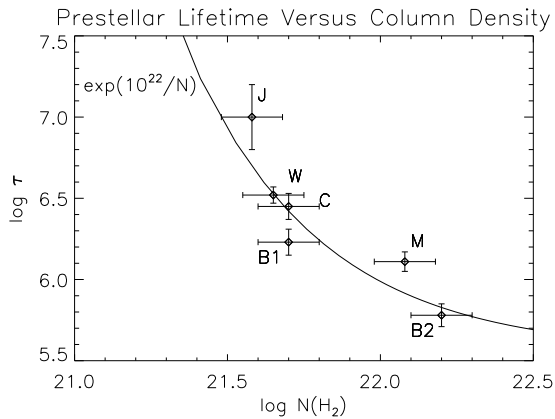}
\caption{Plot of inferred pre-stellar lifetime against column density, 
for each sample discussed in the text.
A best-fit exponential of $\tau \propto e^{(10^{22}/N)}$ is also plotted. 
The reduced chi-squared error on this fit is 1.15.}
\end{center}
\end{figure}

\begin{figure}
\begin{center}
\leavevmode
\epsfxsize=8cm
\epsfbox[0 0 500 350]{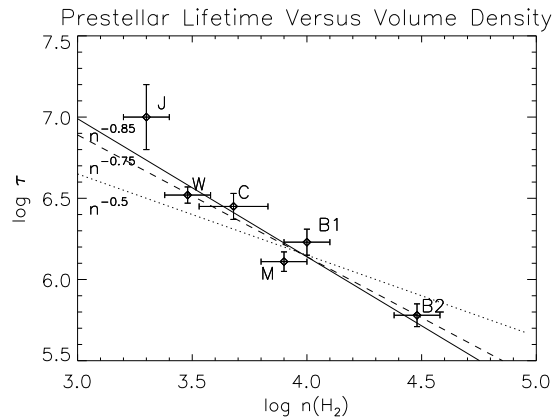}
\caption{Plot of the inferred pre-stellar lifetime against volume density,
for each sample discussed in the text.
A best-fit power-law of $\tau \propto n(H_2)^{-0.85}$
is plotted as a solid line. The reduced chi-squared error on this
fit is 1.0. Functions of $\tau \propto n(H_2)^{-0.5}$ and $\tau \propto
n(H_2)^{-0.75}$ are also shown, as dotted and dashed lines respectively.}
\end{center}
\end{figure}

\begin{figure}
\begin{center}
\leavevmode
\epsfxsize=8cm
\epsfbox[0 0 500 350]{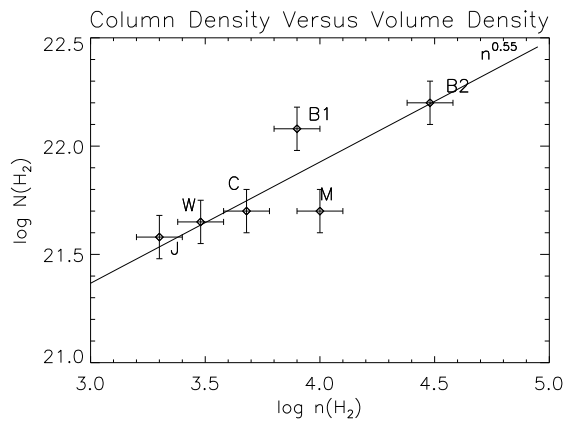}
\caption{Plot of the column density against volume density,
for each sample discussed in the text.
A best-fit power-law of $N(H_2) \propto n(H_2)^{0.55}$
is plotted as a solid line.}
\end{center}
\end{figure}

The fraction of cores with embedded sources has random errors,
and in the catalogues where the number of cores with embedded
sources is large, the 1 $\sigma$ error is simply the square root of 
the number of cores with embedded sources.
The number of cores in the sample presented in this paper has a low
number of cores with embedded sources and therefore the 
uncertainty in the pre-stellar lifetime is dominated by the systematic
effects discussed in section 3.2 above.
These errors are quoted in Table 4.

It is apparent from Table 4 that the percentage of cores with embedded
sources increases with both the mean column density and volume density of 
the cores. Hence the pre-stellar 
core lifetime decreases with both column and
volume densities.
The results are plotted in Figs. 5 \& 6.

Figure 5 shows prestellar lifetime versus column density. Each of the 
points represents one of the data sets listed in Table 4 (B1 signifies
Bourke et al.(1) etc.). We fitted a relation to the data of the form 
$ \tau \propto {\rm exp}(10^{22}/N)$. This is shown as a solid line.
Figure 6 shows pre-stellar lifetime versus volume density (using the same
labelling as in Figure 5). 
Power law fits to the data points were carried out, and the best-fit
was found to be $\tau \propto n(H_2)^{-0.85 \pm 0.15}$,
which is shown as a solid line on Figure 6.

Figure 7 plots column density against volume density for each of the
core data sets in Table 4, using the same labelling once again. The
solid line is a best fit to the data, which is
$N(H_2) \propto n(H_2)^{0.55 \pm 0.13}$. This is
of a similar form to the empirical observations
usually referred to as Larson's
relations (Larson 1981), although with a somewhat different slope.
If we make the assumption that each catalogue is
a representitive subset of all star-forming clouds, and 
the different membership of each subset
is simply due to the characteristic density sensitivity 
of the different tracers used in each study, then 
we can compare these empirical results with theoretical 
predictions.

\subsection{Comparison with theory}

We can compare the dependence of prestellar time-scale on volume density 
seen here, with the ambipolar diffusion time-scale. However, it must be
noted that the densities plotted in Figure 6 are the mean densities for
each core sample, volume-averaged over the whole core in each case.
They are not the central densities of each core. Hence care must be
taken when comparing Figure 6 with ambipolar diffusion models,
which usually plot central density versus time-scale.

The ambipolar diffusion time-scale is proportional
to the ionisation fraction, $\chi_i$.
The ionization of the gas can be caused both by ultra violet radiation and 
cosmic ray ionisation. The canonical form for the relation between
cosmic ray ionisation and volume density is usually taken 
to be a power-law. For example, Mouschovias 
(1991) uses:

\[ \chi_i \propto n(H_2)^{-0.5}, \]

\noindent
which leads to the derivation of the ambipolar diffusion
time-scale relative to density of:

\[ \tau_{AD} \propto n(H_2)^{-0.5}. \] 

\noindent
This behavior is plotted as a dotted line in Fig. 6. When additional 
factors are included (such as chemistry, multiple charge carriers etc.), the
volume density has a slightly different
influence on recombination rates, and 
hence ionisation levels. For example,
McKee (1989) suggests:

\[ \chi_i \propto n(H_2)^{-0.75}, \]

\noindent
leading to:

\[ \tau_{AD} \propto n(H_2)^{-0.75}. \]

\noindent
This is plotted as a dashed line on Figure 6. It can be seen that
the steeper
relation is a closer match to the data, and is in fact consistent
to within 1 $\sigma$.

The fact that the lowest density point on Figure 6 lies
above the best-fit line suggests that at low
densities an additional effect may be starting to become
significant. This is somewhat tentative, but does have a 
possible theoretical explanation. For example,
McKee (1989) treats ionization and recombination in some
detail, and incorporates the role of metals, and UV penetration. 
He derives an expression for the star formation time-scale which 
is dependent on both volume density and column density 
(see his equation 4.5, and his fig. 1). He also shows that UV ionisation 
affects the star formation timescale, yielding an exponential dependence 
on column density (McKee equation 5.2) of the form:

\[ \tau_{SF} \propto e^{(N_c/N)},\]

\noindent
where $\tau_{SF}$ is the star formation timescale and $N_c$ is the critical
column density. $\tau_{SF}$ is not the same as the timescale we have plotted
in Figure 5, since McKee was referring to the timescale in which an entire
molecular cloud is converted to stars and we are considering the timescale
for the dense cores within clouds to form stars.
Hence his timescales are roughly an
order of magnitude greater than ours. However, 
we may be observing a similar exponential behaviour.  
The constant in the exponent, $N_c$, was derived by McKee to be
1.6 $\times$ 10$^{22}$ cm$^{-2}$ (deduced from his critical extinction
estimate of A$_V$=16), which is consistent, to within
the errors, with the value of 1.0 $\times$ 10$^{22}$ cm$^{-2}$ that we
derive in Figure 5. 

The three parameters of column density, volume density and lifetime are
all inter-dependent. Hence Figures 5 \& 6 are not independent. Therefore
the empirical fits which we have applied to these plots are not strictly
separable in the manner we have adopted. However, we used this approach
for the sake of clarity and of demonstrating the potential underlying
physical processes. From these data we cannot make definitive statements
about ionisation mechanisms,
but we have shown that the data, not only from
our current survey, but also from those earlier surveys that we have 
discussed above, are all consistent with a picture in which the
ionisation levels in molecular cloud cores regulate the star formation
timescale.

\section{Conclusions}

We have presented a catalogue of 60 cores situated in  
medium opacity molecular clouds, with the aim of broadening the 
range of physical environments in which star formation has been 
studied. The catalogued cores typically have lower column densities 
and volume densities
than previously studied samples and a 
lower fraction of the cores have formed stars.
We found a clear trend for cloud cores to form stars more rapidly
with increasing volume and column density. We
hypothesised that this can be interpreted in the 
framework of ionisation-regulated star formation. 

\section*{Acknowledgments}

NEJ acknowledges PPARC for studentship funding whilst
at the University of Edinburgh. {\it IRAS} was operated by the US
National Aeronautics and Space Administration. Data handling facilities
for the UK were provided by the Rutherford Appleton Laboratory.
The ISSA images were produced by the Infrared Processing and Analysis
Centre (IPAC) at the Jet Propulsion Laboratory (JPL), California.
The authors would also like to thank STARLINK, and particularly David
Berry, for provision of the COLTEMP routine and assistance in modifying
this routine to produce the core catalogue.

\label{lastpage}

\end{document}